\documentclass[aps,prx,showpacs,twocolumn,superscriptaddress,floatfix]{revtex4-2}

\usepackage{graphicx}
\usepackage{dcolumn}
\usepackage{bm}
\usepackage{amsthm}
\usepackage{xcolor}
\usepackage{braket}
\usepackage{physics}
\usepackage{dsfont}

\usepackage{algorithm}
\usepackage{algpseudocode}
\usepackage{amsmath,amssymb,amsfonts}
\usepackage{braket}
\usepackage{algorithmicx}
\usepackage{algpseudocode}

\newcounter{algorithmctr}
\renewcommand{\thealgorithmctr}{\arabic{algorithmctr}}

\newcommand{\Id}{\mathbb{I}}

\makeatletter
\let\oldState\State
\renewcommand{\State}{\oldState\hangindent=\ALG@thistlm \hangafter=1 }
\makeatother

\usepackage{listings}
\usepackage{tikz}
\usetikzlibrary{quantikz}

\usepackage[caption=false]{subfig}
\usepackage{ragged2e}
\DeclareCaptionJustification{justified}{\justifying}

\usepackage{verbatim}
\usepackage{balance}

\usepackage[colorlinks=true,
            linkcolor=blue,
            citecolor=blue,
            urlcolor=blue]{hyperref}

\usepackage{cleveref}

\usepackage[colorlinks=true,linkcolor=blue,citecolor=blue,urlcolor=blue]{hyperref}

\newtheorem{definition}{Definition}

\def\<{\langle}
\def\>{\rangle}

\begin{document}

\title{Recovery Algorithm for Correlated Errors in Permutation-Invariant Quantum Codes}

\author{Omprakash Chandra}
\email{omprakash.chandra@hdr.mq.edu.au}
\affiliation{School of Mathematical and Physical Sciences, Macquarie University, 2109 NSW, Australia}

\author{Yingkai Ouyang}
\affiliation{School of Mathematical and Physical Sciences, University of Sheffield, Sheffield, S3 7RH, United Kingdom}

\author{Gopikrishnan Muraleedharan}
\affiliation{BTQ Technologies, 16-104 555 Burrard Street, Vancouver, British Columbia, Canada V7X 1M8}

\author{Gavin K. Brennen}
\affiliation{School of Mathematical and Physical Sciences, Macquarie University, 2109 NSW, Australia}
\affiliation{BTQ Technologies, 16-104 555 Burrard Street, Vancouver, British Columbia, Canada V7X 1M8}

\date{\today}

\begin{abstract}
   Quantum Error Recovery (QER) uses knowledge of the error channel acting on a quantum system to find optimal recovery maps. The scheme restores the uncorrupted state with a fidelity exceeding that achieved by noise parameter independent quantum error correction. We use a generic coherent QER map implemented with a quantum circuit acting on the system together with ancillary qubits to recover quantum information stored in permutation invariant (PI) codes. PI codes admit tunable parameters to suit the noise model and benefit from simple recovery operation circuits with reduced addressability requirements, unlike stabilizer codes. We showcase the method by modeling QER in PI codes after collective and local symmetric correlated amplitude-damping (AD) noise, a non-Pauli noise process for which stabilizer codes often require additional overhead. We also propose a new PI code family called CAD codes with explicit examples on 4 and 9 qubits for global symmetric AD errors. We show that CAD9 (supported on 9 qubits) code beats many existing codes by more than one order of magnitude. For the CAD4 code, which perfectly corrects \(1\) global symmetric AD error, the compiled recovery circuit consists of 10 system and system-ancilla gates which can be realized from linear geometric phase gates. Our work provides a direct path from optimized recovery maps to experimentally implementable, low-overhead protocols.

\end{abstract}

\maketitle
\section{Introduction}

Recent breakthroughs~\cite{acharya2025quantum} have significantly reduced the timeline for realizing scalable, digital fault-tolerant quantum computers. Consequently, research into quantum error correction (QEC) has intensified, with a growing emphasis on hardware-efficient strategies. As physical architectures mature, researchers are increasingly looking beyond generic QEC to bespoke, hardware-adapted codes and decoding algorithms. A dominant bottleneck across leading physical platforms, including superconducting circuits, trapped ions, and solid-state spin defects~\cite{krantz2019quantum,bruzewicz2019trapped, kubica2023erasure}—is spontaneous energy decay, mathematically modeled as the amplitude-damping (AD) channel where a qubit transitions from $|1\rangle \rightarrow |0\rangle$ with some probability.

Standard QEC frameworks are mainly based on stabilizer codes, the bedrock of modern fault tolerance. While stabilizer codes are great protecting against Pauli errors, they struggle when it comes to AD errors. AD error is an intrinsically non-Pauli process and stabilizer codes are not naturally built to correct it, often resulting in increased resource overheads when dealing with AD-dominated environments~\cite{jayashankar2022achieving,jackson2016concatenated}. Consequently, developing low-overhead noise-adapted codes and continuous recovery strategies has become a crucial frontier \cite{fletcher2007optimum}. While bespoke AD codes have been proposed to address this \cite{leung1997approximate,Fletcheretal2008, chuang1997bosonic, grassl2018quantum,ouyang2019permutation}, realizing their theoretical potential in the laboratory presents a formidable control challenge. Most existing AD codes require precise, individual qubit addressability for state preparation, syndrome extraction, and control. In densely packed quantum systems, this local addressing is difficult and introduces coherent crosstalk, a major source of operational noise. 

Permutation-invariant (PI) quantum codes offer an elegant alternative~\cite{ouyang2014permutation}. By encoding information in states that remain invariant under any reordering of the qubits, PI codes completely bypass the need for individual addressability. Instead, they require only uniform, collective control over the entire qubit register for QEC \cite{ouyang2026theory} and logical operations \cite{ouyang2025measurement}, which are naturally available in platforms such as cavity-QED architectures. 

Another major source of noise in neutral atoms, photonics, and ion-trap systems is deletion errors, where a qubit is lost and its location is unknown. Efforts have been made to identify the location of the lost qubit and replace it with a fresh qubit. This step is called deletion-to-erasure conversion. The permutational symmetry of the PI codespace naturally makes deletion errors equivalent to erasure. In fact, bespoke PI codes have been constructed to tackle deletion errors \cite{ouyang2021permutation, shibayama2021permutation,aydin2024family, bond2026permutation}. However, AD noise is different. The qubit is still present, but its state has decayed to an unwanted state. Therefore, techniques used to handle deletion errors cannot be directly copied to the AD setting.

This motivates the question of whether one can design PI codes tailored to AD noise? Especially for global symmetric or collective AD noise, where errors preserve the Dicke space to which the PI codespace belongs. In this work, we use this symmetric structure of the errors to introduce collective-amplitude-damping PI codes, or CAD PI codes, designed specifically for global symmetric AD errors. We give explicit examples by constructing the 4-qubit CAD4 and 9-qubit CAD9 PI codes. Note that throughout the paper, global symmetric AD, collective AD, and correlated AD mean the same thing.

Despite these physical advantages, the potential of PI codes remains largely untapped in practical settings. While theoretical bounds on their performance exist, it is not well understood how short-length PI codes actually perform under optimized quantum error recovery (QER), the analog to the discrete syndrome-extraction and Pauli-correction cycles used in stabilizer codes. Furthermore, translating an abstract mathematically optimized recovery map into a realizable set of physical operations using only collective control remains a highly nontrivial challenge. 

We bridge the above gap by studying several short-length PI codes, including \(\mathrm{gnu}\)\cite{ouyang2014permutation}, \(\mathrm{bgm}\)\cite{ouyang2025measurement}, \(\mathrm{bg}\), the 7-qubit Aydin--Alekseyev--Barg (AAB) code \cite{aydin2024family}, the 7-qubit Pollatsek--Ruskai code \cite{pollatsek2004permutationally}, and the 11-qubit Kubischta--Teixeira code \cite{kubischta2024permutation}, along with our newly discovered CAD codes, under optimized QER. We numerically benchmark the codes against global and local symmetric AD noise, showing that optimized QER can suppress the entanglement infidelity of the logical codespace by several orders of magnitude. For example, CAD9 code beats the second best 7-qubit AAB code by more than one order of magnitude at AD strength of \(p=1\times 10^{-3}\) and by more than two order of magnitudes at \(p = 1 \times 10^{-4}\). Asymptotically, CAD codes on \(N\) qubits correct $\sqrt N-1 $ collective AD errors. While, \(\mathrm{gnu}\) codes with parameters $g=n=\sqrt N$ and $u=1$ on $N$ qubits correct $\lfloor \sqrt N-1 \rfloor$ errors, 
which is half of what CAD can correct.

We then turn to another problem: how can these abstract recovery maps be implemented on a near-term quantum device?
We bridge this gap by giving an explicit recipe to implement a recovery channel using ancilla-assisted coherent-control protocol that closely follows the construction introduced by Ticozzi and Viola~\cite{ticozzi2017quantum}. This construction implements a rank-\(K\) quantum channel using \(K-1\) ancilla qubits in \(K-1\) sequential branching steps. Here, the Kraus rank \(K\) denotes the minimum number of Kraus operators required to represent the channel, or equivalently, the rank of its Choi matrix. The approach can be viewed as a coherent simplification of the earlier Lloyd--Viola measurement-and-feedback method~\cite{lloyd2001engineering}, where the intermediate measurements and feedback are replaced by coherent branch storage in ancillas.

Although this coherent CPTP-map construction is general, our focus here is its application to PI codes, where the relevant system dynamics are restricted to the Dicke subspace. In this setting, we give an explicit compilation scheme of the recovery algorithm. The gate complexity scales linearly with the Kraus rank, \(\mathcal{O}(K)\), when counting only the dominant primitives before compilation into physical gates. We then give a compilation of the required system and system--ancilla operations using geometric phase gates (GPGs). As a concrete demonstration, we construct a rank-\(4\) recovery channel corresponding to approximate global symmetric AD noise and compile the required unitaries and controlled unitaries into GPG primitives. As an example of the compilation, we present an explicit recovery circuit for the CAD4 PI code, to recover from first-order truncated global symmetric AD. Finally, we test the complete recovery circuit for the \(((9,1,3))\)-\(\mathrm{bgm}\)-PI code~\cite{ouyang2025measurement}, using both noiseless and noisy GPG sequences. By including a noisy GPG model parametrized by the cooperativity \(C\), we quantify the hardware-level implementation penalty and show that the recovery advantage survives realistic control imperfections.

We stress that our proposal is not intended to replace large-scale fault-tolerant architectures based on surface codes~\cite{fowler2012surface} or quantum LDPC codes~\cite{webster2026pinnacle}. Rather, it should be viewed as a low-overhead, complementary inner layer tailored to AD-dominated noise. With realistic collective-control primitives and only a modest number of qubits, PI codes can substantially suppress the effective AD rate before a larger outer code is applied. One can envision replacing each physical qubit of a higher-level Pauli-based architecture with a small PI-encoded block, applying a noise-adapted recovery at this inner level, and then passing the ``cleaned'' logical qubits to the outer fault-tolerant layer~\cite{kuo2026degenerate}. 

At the same time, our scheme is experimentally attractive in its own right. Since both the code and the recovery are formulated directly in the Dicke subspace, the error information is organized by permutation-symmetric loss sectors rather than by spatially resolved error patterns on individual qubits. This is a major simplification compared with a generic \([[n,k]]\) stabilizer code, where the syndrome space contains \(2^{n-k}\) possible outcomes and a nontrivial classical decoder is required to choose the recovery. Even for state-of-the-art stabilizer decoders, the best achievable decoding complexity is typically linear or near-linear in \(n\), up to slowly growing factors.~\cite{delfosse2021almost}. By contrast, a PI code correcting \(k\) amplitude-damping losses has only \(O(k)\) relevant correctable sectors. In the regimes considered here, where \(k=O(\sqrt{n})\), even a brute-force decoder has only \(O(\sqrt{n})\) possible recovery branches. This reduced decoding structure, together with global control in the Dicke subspace, makes PI codes a particularly natural route towards near-term demonstrations of nontrivial quantum error correction in platforms that do not yet support full fault-tolerant control.

The remainder of the paper develops these results as follows. Sec.~\ref{sec: introduction to quantum error recovery} reviews quantum error recovery, and Sec.~\ref{sec: implementing cptp maps} presents the algorithm for implementing recovery maps as explicit circuits. We then specialize this framework to PI codes: Sec.~\ref{sec:pi-codes} introduces the PI code families studied, while Sec.~\ref{sec: error channels} defines the global and local symmetric AD noise models and their associated recovery maps. Sec.~\ref{sec:PI codes for collective AD channel} introduces new CAD PI codes, explicitly constructed for collective AD noise, followed by Sec.~\ref{sec: detailed implementation of the algorithm for AD noise}, which gives a detailed implementation of the proposed recovery algorithm for collective AD noise. Sec.~\ref{sec: compiling unitaries} decomposes the required recovery operations into Dicke-subspace primitives, and Sec.~\ref{sec:gpg-implementation-elementary-ops} compiles these primitives into GPG sequences. Finally, Sec.~\ref{sec:numerical_results for error recovery} presents numerical results, including an end-to-end noisy-GPG implementation for the \(((9,1,3))\)-bgm code, SDP-optimal recovery benchmarks for short-length PI codes of distance \(3\) including CAD codes, and the correction of spatially correlated AD errors and finally concluding with a discussion in Sec.~\ref{sec:discussion}.

\section{Recovery maps and their implementation} 

\subsection{Quantum error recovery} \label{sec: introduction to quantum error recovery}

We present a high-level overview of the quantum error recovery (QER) problem. 
Given an input quantum state $\rho$ that undergoes a known completely positive 
and trace-preserving (CPTP) noise channel $\mathcal{E}$, the goal of QER is to 
find a recovery CPTP channel $\mathcal{R}$ such that the final state 
$(\mathcal{R} \circ \mathcal{E})(\rho)$ closely approximates the initial state $\rho$
We call the channel \(\mathcal{R}\) a recovery channel for the noisy channel \(\mathcal{E}\). We use the notion of entanglement fidelity \(\mathrm{F_e}\) \cite{PhysRevA.54.2614} to gauge this closeness. The reason can be understood from the following simple argument: If the goal were merely to protect a known density matrix \(\rho\), one could trivially achieve this by tracing out the corrupted output state and re-initializing \(\rho\). While such a ``discard and prepare'' strategy perfectly reproduces the local state \(\rho\), it completely severs any entanglement the state shared with a reference system. Entanglement fidelity detects this loss of quantum correlations, which is why merely recovering the local density matrix does not guarantee \(\mathrm{F_e} = 1\). Therefore, it becomes our choice to evaluate the performance of QER throughout this paper.

In the general setting, input state \(\rho\) is unknown. Therefore, the QER problem would require finding a recovery map that performs well for all input states. This naturally leads to a min--max problem: one maximizes over recovery channels \(\mathcal{R}\), while minimizing over all possible \(\rho\). Intuitively, this means finding the best recovery map for the worst input state. To avoid this more difficult min--max optimization, we follow the average entanglement fidelity approach, where \(\rho\) is chosen to be the maximally mixed state, as in \cite{fletcher2007optimum}. With this choice, we are essentially looking for a recovery map \(\mathcal{R}^* = \operatorname*{argmax}_{\mathcal{R}} \mathrm{F_e}(\rho, \mathcal{R} \circ \mathcal{E})\). Clearly, this is an optimization problem with an objective function linear in \(\mathcal{R}\). This is a special case of the much broader class of convex optimization problems. Having formulated the recovery problem, we now discuss two standard ways of constructing the recovery map: SDP-optimized recovery and the Barnum-Knill/Petz recovery.
 
\subsubsection{Semidefinite optimisation}

We can use semidefinite programming (SDP) to solve the optimization problem mentioned above \cite{fletcher2007optimum}. Note that there exist different possible sets of Kraus operators to represent a quantum channel \(\mathcal{E}\) \cite{nielsen2010quantum}, and similarly for the recovery channel \(\mathcal{R}\) that we are seeking. This means that when searching for the best channel \(\mathcal{R}^*\), we might consider two identical dynamics as different, leading to optimization over redundant degrees of freedom. To counteract this, we use the Choi matrix representation, which is unique for a given quantum channel---a one-to-one correspondence known as the Choi-Jamio\l{}kowski isomorphism \cite{choi1975completely, jamiolkowski1972linear}. In this form, since the CPTP constraints reduce to linear matrix inequalities, the optimization problem can be compactly written as a semidefinite program \cite{audenaert2002optimizing, fletcher2007optimum}. Explicitly, 
it is written as
\begin{equation}
\label{eq:the_optimization_problem}
    \mathcal{X}_{\mathcal{R}}^*
    =
    \operatorname*{argmax}_{\mathcal{X}_{\mathcal{R}}\geq 0}
    \operatorname{Tr}\!\left(
        \mathcal{X}_{\mathcal{R}} C_{\rho,\mathcal{E}}
    \right),
    \qquad
    \operatorname{Tr}_{\mathrm{env}}
    \!\left(
        \mathcal{X}_{\mathcal{R}}
    \right)
    =
    I_{\mathrm{sys}},
\end{equation}
where \(\mathcal{X}_{\mathcal{R}}^*\) denotes the Choi matrix of the recovery map and
\begin{equation}
    C_{\rho,\mathcal{E}}
    =
    \sum_j
    \vert \rho E_j^{\dagger}\rangle\!\rangle
    \langle\!\langle \rho E_j^{\dagger}\vert
\end{equation}
is the matrix in the Choi representation associated with the input state \(\rho\) after the application of the error channel \(\mathcal{E}=\{E_j\}\). Here \(\operatorname{Tr}\) denotes the trace and \(\vert\cdot\rangle\!\rangle\) denotes matrix vectorization. The constraints in Eq.~\eqref{eq:the_optimization_problem} guarantee that the recovery map is physically valid: \(\mathcal{X}_{\mathcal{R}}\geq 0\) enforces complete positivity, while \(\operatorname{Tr}_{\mathrm{env}}(\mathcal{X}_{\mathcal{R}})=I_{\mathrm{sys}}\) enforces trace preservation.

\subsubsection{ Barnum-Knill (or Petz recovery)}

The problem of QER can be solved analytically using the Barnum-Knill (BK) \cite{barnum2002reversing} recovery or Petz recovery scheme \cite{petz1988sufficiency}. The high-level idea is that there exists a recovery channel \(\mathcal{R}^{(p)}\) (superscript \((p)\) simply denotes Petz) that can completely reverse the action of the map \(\mathcal{E}\) by applying a transpose map. The Petz recovery map corresponding to the error map \(\mathcal{E}\) is
\begin{align} \label{eq: petz recovery definition}
   \mathcal{R}^{(p)} := 
\left\{R_i^{(p)} = \rho^{1/2} E_i^{\dagger} \, \mathcal{E}(\rho)^{-1/2} \right\},
\end{align} 
where subscript \(i\) denotes recovery operator \(R_{i}^{(p)}\) corresponding to Kraus operator \(E_i\) of the channel \(\mathcal{E}\). Note that the channel \(\mathcal{R}^{(p)}\) is a completely positive and trace-non-increasing map. If $\mathcal E(\rho)$ is not a full rank operator, then $\mathcal E(\rho)^{-1/2} = \tau^{1/2}$, where $\tau$ is the pseudo-inverse or Moore-Penrose inverse of $\mathcal E(\rho)$. Applied to the context of QEC, we choose $\rho$ as the maximally mixed state on the codespace. It is worth emphasizing that if the Kraus operators \(E_i\) satisfy the Knill-Laflamme error correction (KL-ECC) conditions \cite{PhysRevA.55.900}, the recovery map perfectly corrects the errors, ensuring \(\mathcal{R}^{(p)}\circ \mathcal{E}(\rho) = \rho\) while achieving an optimal entanglement fidelity of \(\mathrm{F_e}(\rho, \mathcal{R}^{(p)}\circ \mathcal{E})=1\). However, even if the KL-ECC conditions are not satisfied, it might still be possible to perfectly recover a specific input density matrix such that \(\mathcal{R}^{(p)}\circ \mathcal{E}(\rho) = \rho\), but fail to achieve an entanglement fidelity of \(1\). This distinction arises because entanglement fidelity as mentioned before is a much more stringent, the gold standard for QEC. 

Equipped with these different recovery methods, we now turn to the main question of this work: how do we actually implement these recovery maps? The maps discussed above give mathematical CPTP channels, but do not directly prescribe a physical recovery protocol. We address this question in the following section.

\subsection{Implementing Recovery maps} 
\label{sec: implementing cptp maps}
\begin{figure*}[t!]
    \centering
    \includegraphics[
        width=0.9\textwidth,
        height=0.35\textheight,
        keepaspectratio,
        trim={0.05cm 0.15cm 0.15cm 0.1cm},
        clip
    ]{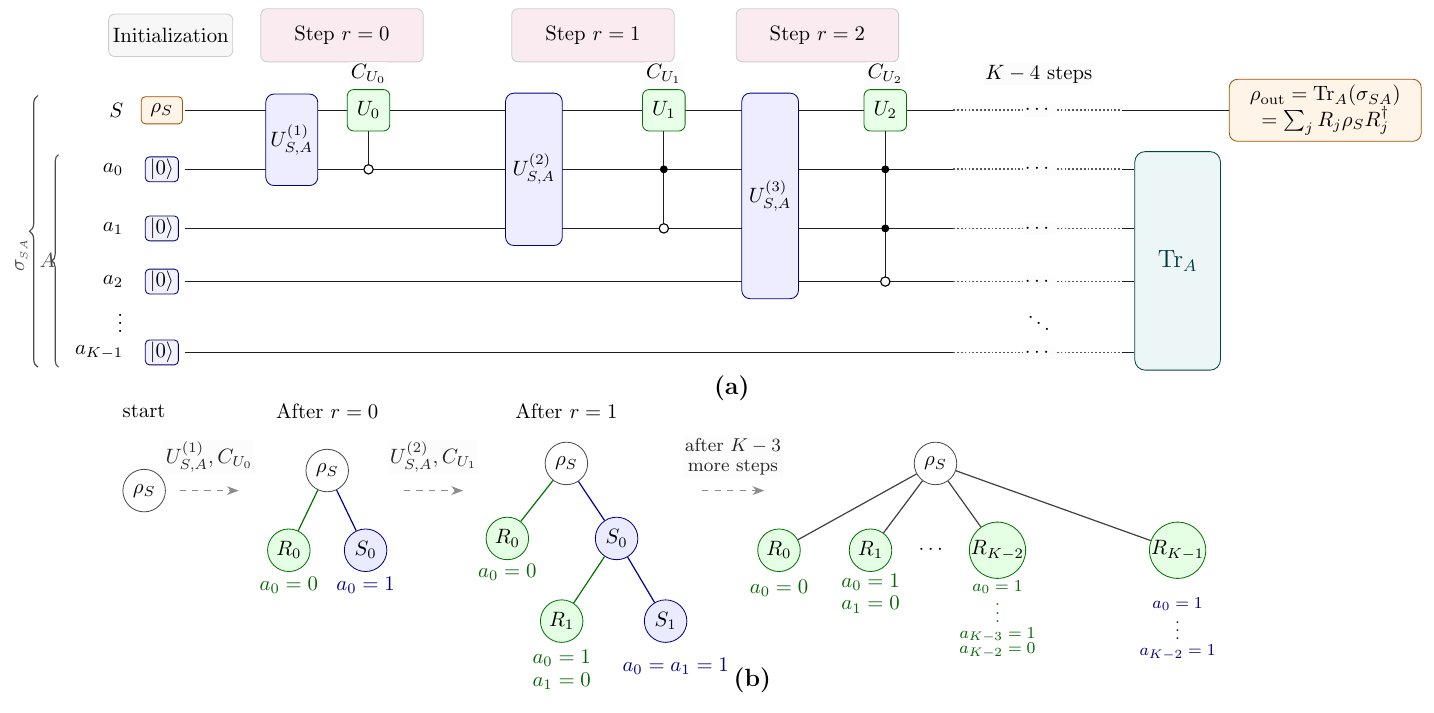}
    \vspace{-2mm}
    \caption{High-level overview of Algorithm~\ref{algo:cptp-implementation}. Here \(\rho_S\) is the system on which we want to implement a rank-K CPTP map \(\mathcal{R} = \{R_j\}_{j=0}^{K-1}\). While \(A\) represent \(K-1\) ancillas in the all-zero state.
(a) The circuit consists of \(K-1\) sequential branching steps. At each step, a system--ancilla unitary \(U^{(r)}_{S,A}\) and the corresponding feedback unitary \(C_{U_r}\) separates one Kraus branch from the remaining branch. After the final step, tracing out the ancilla \(A\) realizes the output state \(\rho_{\mathrm{out}}=\mathrm{Tr}_A(\sigma_{SA})=\sum_{j=0} ^{K-1} R_j\rho_S R_j^\dagger\).
(b) Tree representation of the same coherent branching process. Each step resolves one Kraus branch \(R_r\) (in green), while the unresolved branch \(S_r\) (in blue) is passed to the next stage. After \(K-1\) steps, all Kraus branches \(R_0,\ldots,R_{K-1}\) have been coherently generated and stored in the ancilla branch labels.}
    \label{fig:cptp_cartoon}
    \vspace{-3mm}
\end{figure*}
We use Algorithm.~\ref{algo:cptp-implementation} to implement an arbitrary finite Kraus-rank CPTP map, which we later apply to recovery maps. The algorithm follows the coherent implementation scheme introduced by Ticozzi and Viola~\cite{ticozzi2017quantum}. This scheme builds on the Lloyd--Viola measurement-and-feedback method~\cite{lloyd2001engineering}, where the system is coupled to an ancilla, the ancilla is measured, and a feedback unitary is applied conditioned on the measurement outcome. In the coherent version, the binary branching is implemented without intermediate measurements: the ancillas store the branch information, the corresponding polar unitaries of the Kraus operators are applied as controlled feedback, and tracing out the ancillas realizes the desired CPTP map. Unlike the Lloyd--Viola setting, where Trotterized sequences are used to engineer continuous-time open-system dynamics, the Ticozzi--Viola protocol directly implements a finite Kraus-rank CPTP map and therefore does not require Trotterization. In this work, we use this protocol as the starting point for compiling optimized recovery maps, and later specialize it to recovery channels acting on the PI/Dicke subspace. Fig.~\ref{fig:cptp_cartoon} presents a high-level overview of the algorithm.
\begin{center}
\refstepcounter{algorithmctr}
\label{algo:cptp-implementation}

\hrule height 0.9pt
\vspace{0.2mm}

\noindent\textbf{Algorithm \thealgorithmctr} \hspace{0.1cm}
\textsc{ImplementCPTP}\((\rho_S,\{R_r\}_{r=0}^{K-1})\)

\vspace{0.2mm}
\hrule
\vspace{0.2mm}

\begingroup
\fontsize{8}{6.8}\selectfont
\setlength{\baselineskip}{10pt}
\setlength{\abovedisplayskip}{0pt}
\setlength{\belowdisplayskip}{0pt}
\setlength{\abovedisplayshortskip}{0pt}
\setlength{\belowdisplayshortskip}{0pt}
\setlength{\jot}{0pt}
\setlength{\parskip}{0pt}
\renewcommand{\arraystretch}{0.72}

\begin{algorithmic}[1]

\State \textbf{Input:} \(\rho_S\), K Kraus operators \(\{R_r\}_{r=0}^{K-1}\).
\State \textbf{Resources:} Ancilla register \(A=\{a_0,a_1,\cdots, a_{K-2}\}\), initialized in \(\ket{0}_A=\ket{0}^{\otimes(K-1)}\).
\State \textbf{Initialization:} \(\sigma_{SA} = \rho_S\otimes\ket{0}_A\bra{0}_A\).

\State \textbf{Polar decomposition:}
\For{\(r=0\) to \(K-1\)}
    \State Compute \(B_r\leftarrow(R_r^\dagger R_r)^{1/2}\) and \(U_r \leftarrow R_rB_r^{-1}\) such that \(R_r=U_rB_r\).
\EndFor

\State \textbf{Main Loop:}
\For{\(r=0\) to \(K-2\)}
    \If{\(r=0\)}
        \State Define \(S_0=\left(\sum_{j=1}^{K-1}B_j^2\right)^{1/2}\).
        \State  Compute the system--ancilla (S, \(a_0\)) unitary
        \(U_{S,a_0}^{(1)}\leftarrow\begin{pmatrix} B_0 & S_0\\ S_0 & -B_0\end{pmatrix}.\)
        \State Set \(\mathbf{U}_{S,A}^{(1)}=U_{S,a_0}^{(1)} \otimes \Id_{\mathbf{A}\setminus\{a_0\}}\),
        \State Apply
        \(
        \sigma_{S,A}\leftarrow
        \mathbf{U}_{S,A}^{(1)}
        \sigma_{S,A}
        \mathbf{U}_{S,A}^{(1)\dagger}.
        \)
        \State Define \(P^{(r=0)}=\ket{0}\bra{0}_{a_0}\) and
        \(
        C_{U_0}^{(a_0)}
        =
        \Id_{S,A}
        +
        (U_0-\Id_S)\otimes P^{(r=0)}.
        \)
        \State Apply
        \(
        \sigma_{S,A}\leftarrow
        C_{U_0}^{(a_0)}
        \sigma_{S,A}
        C_{U_0}^{(a_0)\dagger}
        \). 

    \Else
        \State Define
        \(
        S_r=\left(\sum_{j=r+1}^{K-1}B_j^2\right)^{1/2},
        \ 
        \widetilde{B}_r=B_rS_{r-1}^{+},
        \ 
        \widetilde{S}_r=S_rS_{r-1}^{+}.
        \)
        \State Compute the system--ancilla unitary
        \(
        U_{S,a_r}^{(r+1)}
        \leftarrow
        \begin{pmatrix}
        \widetilde{B}_r & *\\
        \widetilde{S}_r & *
        \end{pmatrix},
        \)
        \Comment{where \(*\) denotes a unitary extension.}
        \State Define
        \(
        Q^{(r-1)}
        =
        \bigotimes_{\ell=0}^{r-1}
        \ket{1}\bra{1}_{a_\ell}
        \)
        and
        \(
        \mathbf{U}_{S,A}^{(r+1)}
        =
        \Id_{S,A}
        +
        Q^{(r-1)}
        \otimes
        \left(
        U_{S,a_r}^{(r+1)}
        -
        \Id_{S,a_r}
        \right),
        \)
        \Comment{with identities on unused ancillas implicit}.
        \State Apply
        \(
        \sigma_{S,A}\leftarrow
        \mathbf{U}_{S,A}^{(r+1)}
        \sigma_{S,A}
        \mathbf{U}_{S,A}^{(r+1)\dagger}.
        \)
        \State Define
        \(
        P^{(r)}
        =
        \left(
        \bigotimes_{\ell=0}^{r-1}
        \ket{1}\bra{1}_{a_\ell}
        \right)
        \otimes
        \ket{0}\bra{0}_{a_r}
        \)
        and
        \(
        C_{U_r}^{(a_r)}
        =
        \Id_{S,A}
        +
        (U_r-\Id_S)\otimes P^{(r)}.
        \)
        \State Apply
        \(
        \sigma_{S,A}\leftarrow
        C_{U_r}^{(a_r)}
        \sigma_{S,A}
        C_{U_r}^{(a_r)\dagger}.
        \)
    \EndIf
\EndFor

\State \textbf{Final feedback:}
\State Define \(P^{(K-1)}
=
\bigotimes_{\ell=0}^{K-2}
\ket{1}\bra{1}_{a_\ell}\).
\State Define \(C_{U_{K-1}}^{(a_{K-2})}
=
\Id_{S,A}
+
(U_{K-1}-\Id_S)\otimes P^{(K-1)}\).
\State Apply \(\sigma_{S,A}\leftarrow
C_{U_{K-1}}^{(a_{K-2})}
\sigma_{S,A}
C_{U_{K-1}}^{(a_{K-2})\dagger}\).

\State \Return \(\rho_{\mathrm{out}}=\operatorname{Tr}_{A}(\sigma_{S,A})\).

\end{algorithmic}
\endgroup

\vspace{0.2mm}
\hrule
\end{center}
Here \(S_{r-1}^{+}\) denotes the Moore--Penrose pseudo-inverse of \(S_{r-1}\). The block forms of \(U_{S,a_0}^{(1)}\) and \(U_{S,a_r}^{(r+1)}\) fix only the first block column; the remaining blocks, denoted by \(*\), are chosen to complete a unitary. This is possible because each first block column defines an isometry on the active support. The feedbacks \(C_{U_r}^{(a_r)}\) apply the polar unitary \(U_r\) only on the newly separated Kraus branch, while the remainder branch is propagated to the next split. 

\paragraph*{Gate complexity.}
For a rank-\(K\) CPTP map, the protocol requires \(K-1\) ancilla qubits and \(K-1\) binary splitting stages. We would like to reinstate that the algorithm can be used to implement arbitrary CPTP map on a finite-dimensional system. However, in this paper, we focus on PI codes, where the physical system is restricted to the Dicke subspace. Therefore, the relevant gate-complexity estimates are stated in terms of Dicke-subspace primitives. Counting only the dominant coherent operations, namely system-only Dicke-space unitaries and controlled system--ancilla Dicke-space unitaries, the protocol complexity scales linearly with the Kraus rank, \(O(K)\). After compiling these Dicke-space primitives into geometric phase gate sequences (GPG), the GPG-level cost scales as \( n_{\mathrm{GPG}}=O(KN^2)\).
For a generic CPTP map on the Dicke subspace, \(K\leq (N+1)^2\), giving the worst-case scaling \(O(N^4)\); for fixed-rank recovery maps, the cost is \(O(N^2)\). We also note that the same algorithm can be implemented using \(\mathcal{O}(\log{K})\) ancilla qubits. However, we chose a unary encoding (representing the integer $0\leq k \leq K$ via a contiguous block of ones, \(\ket{ 1^k 0^{K-k}}\)) to keep the implementation elegant and simple. Following the unary iteration framework introduced by Babbush et al.~\cite{Babbush_2018}, this linear-qubit mapping eliminates the heavy multi-controlled gate overhead typical of binary registers, allowing for a much cleaner, linear-depth circuit design. See Appendix~\ref{app:gate-complexity of algorithm 1} for a detailed gate count analysis. We discuss the compilation of these unitaries into GPG primitives in detail in Sec.~\ref{sec: compiling unitaries}.

As mentioned earlier, the proposed method applies to any finite-rank CPTP map. The remaining question is when this implementation becomes physically useful and efficient. We now address this question for PI codes, with the following section serving as a concrete example of the algorithm. 

\section{Recovering Permutation Invariant Codes}

\subsection{Permutation invariant (PI) codes}\label{sec:pi-codes}

Stabilizer codes are the most widely used framework for encoding quantum information because they provide a clean theory of error correction. However, their implementation requires individual addressability and control of the physical qubits, which remains challenging in many hardware platforms. A natural question to ask is: is there a non-trivial quantum code that can be implemented without requiring individual addressability and control of every physical qubit? One answer is provided by PI codes. Owing to their permutational symmetry, PI codes require only global addressability and control for error correction and logical computation, which is natively available in most hardware platforms.

The idea of encoding quantum information into a permutation-invariant codespace dates back to early work by Ruskai~\cite{ruskai2000pauli}. The key motivation is that in many physical hardware platforms, Pauli exchange interactions (\(\propto S_i\cdot S_j\)) naturally arise as a first-order effect of couplings between particles, leading to exchange or swap-type errors. Encoding information into a subspace that remains invariant under such interactions, a PI codespace, naturally eliminates these errors.

Unlike stabilizer codes~\cite{gottesman1997stabilizer}, PI codes are generally not defined as simultaneous +1 eigenspaces of a commuting Pauli stabilizer group. As a consequence, the stabilizer based syndrome extraction for error correction does not directly apply. Error correction for PI codes instead follows a separate theory tailored to permutation-invariant code spaces~\cite{ouyang2026theory}. PI codes offer useful advantages from the viewpoint of quantum computation. For example the native transversal implementation of certain non-Clifford gates, motivating both code-switching protocols between stabilizer and PI codes~\cite{ouyang2025measurement} and magic-state distillation protocols directly on PI codes~\cite{leitch2026magic} to lower the overhead associated with non-Clifford gates. 

Furthermore, the permutational symmetry of PI codes makes them well suited to deletion-error models, where the location of a lost qubit is not known~\cite{hagiwara2020four,ouyang2021permutation,shibayama2021permutation}. Beyond error correction, PI codes have also found applications in quantum storage~\cite{ouyang2021quantum}, quantum communication~\cite{ouyang2019permutation}, and quantum sensing~\cite{ouyang2021robust,ouyang2022quantum}, highlighting their versatility across different quantum-information tasks.


The $(N+1)$-dimensional PI codespace is spanned by the Dicke states, defined as the equal superposition of all $N$-qubit computational basis states with Hamming weight $w$, namely $\ket{D_w^{\,N}} = \binom{N}{w}^{-1/2} \sum_{\mathrm{wt}(x)=w} \ket{x}$. These states reside in the maximum total angular momentum subspace ($J=N/2$) of the $N$ qubits, acting as simultaneous eigenstates satisfying $J^2 \ket{D_w^{\,N}} = \frac{N}{2}(\frac{N}{2}+1)\ket{D_w^{\,N}}$ and $J_z \ket{D_w^{\,N}} = (\frac{N}{2}-w)\ket{D_w^{\,N}}$. Here \(J_\alpha = (\Sigma_{i=1} ^N \sigma_\alpha ^{(i)})/2\) for \(\alpha = x,y,z\) denotes the collective spin operator where \(\sigma_{\alpha}^{(i)}\) is the usual Pauli operator \(\sigma_{\alpha}\) acting on the (i)-th qubit and identity on all other qubits. The total angular momentum operator is \(J^2 = J_x^2 + J_y^2 + J_z^2\). Note that for convenience we omit hats from operators throughout this paper.  

A general PI code with single logical qubit is defined by two orthonormal logical states constructed entirely from this basis: $\ket{0}_L = \sum_{i=0}^{N} a_i \ket{D^N_i}$ and $\ket{1}_L = \sum_{i=0}^{N} b_i \ket{D^N_i}$, where $a_i,b_i \in \mathbb{C}$ subject to the orthogonality condition $\sum_{i=0}^{N} a_i^*\, b_i = 0$. While PI codes can encode multiple logical qubits \cite{Ouyang2016, Kubischta2025}, we restrict our attention to this two-dimensional codespace. With this framework established, we now define the specific PI code families, including the \(\mathrm{gnu, bgm}\) and 7-qubit PI codes, that we will analyze throughout this work.

\begin{definition} \label{def: gnu codes}
    \(\mathrm{gnu}\) codes: Logical basis states of gnu-codes are defined as \cite{ouyang2014permutation}, 
\begin{equation}
\begin{aligned}
\ket{0_L} &= \sum_{\substack{\ell~\mathrm{even}\\ 0\le \ell \le n}} \sqrt{\frac{\binom{n}{\ell}}{2^{\,n-1}}}\; \ket{D^{gnu}_{g\ell}}, \\ 
\ket{1_L} &= \sum_{\substack{\ell~\mathrm{odd}\\ 0\le \ell \le n}} \sqrt{\frac{\binom{n}{\ell}}{2^{\,n-1}}}\; \ket{D^{gnu}_{g\ell}},
\end{aligned}
\end{equation}
where \(g, n, \text{and} \  u  \geq 1\) are positive integers , and  \(N = gnu\) is the total number of physical qubits.
\end{definition} If \(g=n=2t+1\) and \( u \geq 1\), gnu codes correct arbitrary \(t\) qubit errors. While if \(g=t+1, n > 3t \) and \( u \geq 1 + \frac{t}{gn}\), the gnu code corrects \(t\) spontaneous decay errors. In the large-$n$ limit, the $gnu$ codes can be viewed as approaching the structure of a GKP code \cite{gottesman2001encoding}. 
In fact, in appropriate parameter regimes, $gnu$-codes may be understood as a discretized analogue of corresponding limiting forms of GKP codes.     

\begin{definition} \label{def: bg codes}
    \(\mathrm{bg}\) codes: Logical basis states of b,g-codes are
\begin{equation}
\begin{aligned}
|0_{b,g}\rangle_L &:= \frac{\big(\sqrt{2b - g}|D_{0}^{2b+g}\rangle\big) + \sqrt{2b + g}|D_{2b}^{2b+g}\rangle}{\sqrt{4b}}, \\
|1_{b,g}\rangle_L &:= \frac{\big(\sqrt{2b - g}|D_{2b+g}^{2b+g}\rangle\big) + \sqrt{2b + g}|D_{g}^{2b+g}\rangle}{\sqrt{4b}},
\end{aligned}
\end{equation}
where \(g\) is a positive integer, \(2b \geq g+1\) and number of physical qubits is \(N=2b+g\). 
\end{definition}
The interesting aspect of the \(b,g\) code is the transversal application of the gate \(Z (\pi/b)^{\otimes N}\) applies a logical-Z rotations with angle \(\pi g/b\). By carefully choosing \(b\) and \(g\), we can apply logical-Z rotation with angles that lie outside the Clifford hierarchy. For example, \(b=4, g=1\) gives us transversal logical-Z rotation of angle \(\pi/4\) which is the popular \(T\) gate. Ability to do these gates transversally opens up attractive options of finite gate set for universal quantum computation as opposed to the traditional \(\{H,S,CNOT\} \cup \{T\}\).

\begin{definition} \label{def: bgm codes}
\(\mathrm{bgm}\) codes: The logical basis states of the \((b,g,m)\) codes are defined as \cite{ouyang2025measurement}

\begin{equation}
\begin{aligned}  
\ket{0_{b,g,m}}  
&= \sum_{k=0}^{m} \sqrt{\binom{m}{k}}  
\,\frac{\gamma_{b,g,m,k}\ket{D^{N}_{2kb}}}{2^{m}\sqrt{(2m-1)!!}},  \\  
\ket{1_{b,g,m}} &= X^{\otimes N}\ket{0_{b,g,m}},  \\  
\text{where } \  \gamma_{b,g,m,k}  
&= b^{-m/2} \left( \prod_{i=k+1}^{m}\sqrt{2ib-g} \right) \\
&\quad \times \left( \prod_{j=m-k+1}^{m}\sqrt{2jb+g} \right), \\
&\quad \text{for } k = 0,\ldots,m.  
\end{aligned}
\end{equation}

and \(N = 2bm+g\). 
\end{definition}

Note that the \((b,g,1)\) reduces to the \((b,g)\) code. 
When \(g,\,2b-g \ge 2t+1\) and \(m \ge t\), the distance is at least \(2t+1\), and hence corrects at least \(t\) errors. Similar to \(\mathrm{bg}\) code, the parameters of \(\mathrm{bgm}\) code can be tuned to support transversal logical \(Z(\pi g/b)\) rotations with the additional \(m\) parameter to achieve higher distances.

\begin{definition} \label{def: 7 qubit code}
   7 qubit AAB code: The Aydin--Alekseyev--Barg (AAB) code was introduced in Example 4 of the paper \cite{aydin2024family}. The logical codewords are 
    \begin{align}
        \ket{0}_L & := \sqrt{3/10} \ket{D^7_0} + \sqrt{7/10} \ket{D^7_5}, \\ 
        \ket{1}_L & := \sqrt{7/10} \ket{D^7_2} - \sqrt{3/10} \ket{D^7_7}. \nonumber
    \end{align}
\end{definition}
The code has distance \(3\) and therefore can correct one arbitrary error, \(((7,1,3))\).

\begin{definition} \label{def:7_qubit_pollatsek_ruskai}
7-qubit PR code: Pollatsek-Ruskai code \cite{pollatsek2004permutationally} is parameterized by $\epsilon \in \{\pm 1\}$, with its logical basis states defined as:
\begin{align}
\ket{0}_L (\epsilon) 
&= \frac{1}{8} \Big( \epsilon\sqrt{15}\ket{D^7_0} - \sqrt{7}\ket{D^7_2} \\ \nonumber 
&\qquad + \epsilon\sqrt{21}\ket{D^7_4} + \sqrt{21}\ket{D^7_6} \Big), \\ \nonumber
\ket{1}_L (\epsilon) 
&= \frac{1}{8} \Big( \sqrt{21}\ket{D^7_1} + \epsilon\sqrt{21}\ket{D^7_3} \\ \nonumber
&\qquad - \sqrt{7}\ket{D^7_5} + \epsilon\sqrt{15}\ket{D^7_7} \Big).
\end{align}
\end{definition}

For either choice of $\epsilon$, the construction yields a valid code with a distance of $3$, enabling it to correct any arbitrary single-qubit error.

\begin{definition} \label{def: 11 qubit code}
11-qubit KT code: The logical codewords of Kubischta--Teixeira code are \cite{kubischta2024permutation} 
\begin{align}
\ket{0}_L:
&=
\frac{1}{4}
\left(
\sqrt{5}\ket{D^{11}_{0}}
+
\sqrt{11}\ket{D^{11}_{8}}
\right),
\\ \nonumber
\ket{1}_L:
&=
\frac{1}{4}
\left(
\sqrt{11}\ket{D^{11}_{3}}
+
\sqrt{5}\ket{D^{11}_{11}}
\right).
\end{align}
\end{definition}
The code has distance \(3\), and supports a
transversal implementation of the logical \(T=Z(\pi/4)\) gate. Note that this code is same as \(\mathrm{b=4,g=3}\) instance of \(\mathrm{bg}\) code given in def.~\ref{def: bg codes}. 

\begin{table}[h!]
    \centering
    \begin{ruledtabular}
        \begin{tabular}{ccc}
        \textbf{Code} & \textbf{Code properties} & \textbf{Parameters} \\ 
        \colrule
        7-qubit AAB & \(((7,1,3))\) & Def.~\ref{def: 7 qubit code} \\
        7-qubit PR & \(((7,1,3))\) & Def.~\ref{def:7_qubit_pollatsek_ruskai} \\ 
        \(\mathrm{gnu}/\mathrm{bg}/\mathrm{bgm}\) & \(((9,1,3))\) & Def.~\ref{def: gnu codes},\ref{def: bg codes}, \ref{def: bgm codes} \\
        11-qubit KT & \(((11,1,3))\) & Def.~\ref{def: 11 qubit code} \\ 
        \end{tabular}
    \end{ruledtabular}
    \caption{
    List of short-length PI code instances used for numerical benchmarking, derived from the definitions in Sec.~\ref{sec:pi-codes}. The \(((9,1,3))\) code is a special case of the \(\mathrm{gnu}\), \(\mathrm{bg}\), and \(\mathrm{bgm}\) constructions, obtained respectively with \(g=n=3,\ u = 1\), \(b=3,\ g=3\), and \(b=3,\ g=3,\ m=1\).    }
    \label{tab:pi-codes-studied}
\end{table}
 
\subsection{Error Channels}
\label{sec: error channels}

A quantum system is inevitably coupled to its surrounding environment, which can appear through stray magnetic or electric fields, phonon excitations, thermal fluctuations, or other uncontrolled degrees of freedom. A general system--environment interaction can be written as
\begin{equation} 
H_{\mathrm{int}} \propto
\sum_{j=1}^{N}\sum_{\alpha}\sum_{i}
S_i^{(j)} \otimes B_{\alpha i}(r_j),
\label{eq:interaction hamiltonian}
\end{equation}
where \(S_i^{(j)}\) denotes the \(i\)-th single-spin operator acting on spin \(j\), with
\(
S_i^{(j)} \in 
\left\{
\sigma_x^{(j)},\sigma_y^{(j)},\sigma_z^{(j)},
\sigma_+^{(j)},\sigma_-^{(j)}
\right\}
\). 
Here, \(B_{\alpha i}(r_j)\) denotes the corresponding bath operator for coupling channel \(i\) and environment mode \(\alpha\), evaluated at the spin position \(r_j\). The nature of the resulting noise depends heavily on how the spacing between the spins compares to the wavelength \(\lambda\) of the relevant environment mode. A stray field imparts a phase \(e^{ik \cdot r_j}\), where \(k = 2\pi/\lambda\). If the separation between spins is much smaller than the wavelength of the mode (i.e., \(\vert r_i - r_j \vert \ll \lambda\)), the phase factors become identical (\(e^{ik \cdot r_i} \approx e^{ik \cdot r_j}\)). In this regime, the system interacts with the environment collectively, which we model as global symmetric noise. Conversely, if the environment can resolve individual spins (\(\vert r_i - r_j \vert \gtrsim \lambda\)), the errors act independently and the noise is referred to as local and furthermore if the strength of errors on each qubit is identical we refer to it as local symmetric noise. 

A dominant source of error in platforms such as superconducting qubits and photonic systems is spontaneous energy decay, which is modeled by an amplitude-damping (AD) channel. AD is a non-Pauli error channel, and generic stabilizer-code approaches do not directly exploit its structure; more broadly, handling related non-Pauli processes such as leakage, loss, or erasure often requires additional detection, ancilla, or syndrome-extraction resources~\cite{chandra2025nonlocal,chow2024circuit}. This motivates codes tailored specifically to AD noise~\cite{leung1997approximate,chuang1997bosonic}. We show in Section~\ref{sec:PI codes for collective AD channel} that bespoke PI codes for AD errors can be designed using just four physical qubits, and that the corresponding recovery circuit can be easily constructed using our proposed algorithm. With this motivation, we study the effect of both global and local symmetric AD noise on PI codes. The Lindbladian for global symmetric AD noise can be written as
\begin{equation} \label{eq:global AD}
    \mathcal{L}_{\text{global AD}}(\rho) = \gamma_- \left( J_- \rho J_+ - \frac{1}{2} \{ J_+ J_-, \rho \} \right),
\end{equation}
where $J_- = J_x - iJ_y$ and $J_+ = J_-^{\dagger}$ denote the collective lowering and raising operators, respectively, and $\gamma_-$ is the decay rate. Since the errors are polynomial in \(J_-\), they generate multi-qubit decay events or spatially correlated AD channel.

The Lindbladian for local symmetric AD is given by:
\begin{equation} \label{eq: local AD}
    \mathcal{L}_{\text{local AD}}(\rho) = \gamma_- \sum_{i=1}^N \left( \sigma_-^{(i)} \rho \sigma_+^{(i)} - \frac{1}{2} \{ \sigma_+^{(i)} \sigma_-^{(i)}, \rho \} \right), 
\end{equation}

where $\sigma_-^{(i)} = \sigma_x^{(i)} - i\sigma_y^{(i)}$ and $\sigma_+^{(i)} = (\sigma_-^{(i)})^{\dagger}$ are the local lowering and raising operators for the $i$-th qubit, and $\gamma_-$ represents the decay rate. Since the error terms act independently on each qubit, the noise is independent and identically distributed (i.i.d.).

We derive a closed-form expression for the channel generated by the global symmetric AD Lindbladian in Eq.~\eqref{eq:global AD}(see Appendix.~\ref{appendix: closed form derivation of AD channel in doubled-space} for the full analytical derivation). To construct practical recovery circuits, we approximate the resulting dynamics up to second order in time (t). Our numerical results show that this second-order truncation accurately captures the relevant behavior, with only a negligible loss in performance. For the case of second order truncation, we obtain following explicit set of Kraus operators:
\begin{align} \label{app:final set of kraus operators for truncated dynamics}
    K_0 &= \mathbb{I} -\frac{\gamma_{-} t}{2} J_+J_- +\frac{\gamma_-^2 t^2}{8} (J_+J_-)^2, \nonumber \\ 
    K_1 &= \sqrt{\gamma_- t}\, J_- -\frac{\gamma_-^{3/2} t^{3/2}}{4} \Bigl( J_-(J_+J_-) + (J_+J_-)J_- \Bigr), \nonumber \\
    K_2 &= \frac{\gamma_- t}{\sqrt{2}}\, J_-^2, \nonumber \\ 
    K_3 &= \sqrt{\mathbb{I} - K_0^{\dagger} K_0 - K_1^{\dagger}K_1 -K_2^{\dagger}K_2}.
\end{align}
The derivation is provided in Appendix.~\ref{app:truncated global symmetric AD}. 

The corresponding recovery map $\mathcal{R}$ for this channel, obtained via the near-optimal Petz recovery scheme~\eqref{eq: petz recovery definition}, is given by the recovery Kraus operators

\begin{align} \label{eq: recovery kraus operators for AD channel}
R_0 &= \rho^{1/2}
\left[
\mathbb{I} - \frac{\gamma_- t}{2} J_+ J_-
+ \frac{\gamma_-^2 t^2}{8} (J_+ J_-)^2
\right]
\nonumber\\
&\quad \times \mathcal{E}(\rho)^{-1/2},
\nonumber\\[0.5em]
R_1 &= \rho^{1/2}
\left[
\sqrt{\gamma_- t} J_+
- \frac{\gamma_-^{3/2} t^{3/2}}{4}
\Big(
J_+ J_- J_+ + J_+ J_+ J_-
\Big)
\right]
\nonumber\\
&\quad \times \mathcal{E}(\rho)^{-1/2},
\nonumber\\[0.5em]
R_2 &= \rho^{1/2}
\left[
\frac{\gamma_- t}{\sqrt{2}} J_+^2
\right]
\mathcal{E}(\rho)^{-1/2},
\nonumber\\[0.5em]
R_3 &= \rho^{1/2}
\left[
\sqrt{
\mathbb{I} - K_0^\dagger K_0
- K_1^\dagger K_1
- K_2^\dagger K_2
}
\right]
\nonumber\\
&\quad \times \mathcal{E}(\rho)^{-1/2}.
\end{align}
For an explicit derivation, see Appendix~\ref{appendix: sec: petz_recovery_second_order}.

\subsection{PI codes for collective AD channel} \label{sec:PI codes for collective AD channel}

In this subsection, we explore whether new PI codes can be designed specifically for the collective AD channel in Eq.~\eqref{eq:global AD}. The errors of this channel are polynomials in the collective lowering operator $J_-$. A code that corrects $k$ AD errors will perfectly correct all errors that are polynomials in $J_-$ of degree at most $k$. We use the Knill-Laflamme \cite{PhysRevA.55.900} quantum error correction criterion, and the linear algebra technique in \cite{ouyang2019permutation,movassagh2020constructing} to design a PI code that encodes a single logical qubit and corrects $k$ collective AD errors.
 
For this technique, we choose the code to be gapped in the Dicke basis just like the gnu codes defined in Eq.~\eqref{def: gnu codes}, but do not fix the amplitudes a priori. We solve for the amplitudes in the logical codewords.
We define the logical codewords as 
\begin{align}
    |0_L\> =\sum_{j\ {\rm even}}
    c_j |D^{(k+1)n}_{(k+1)j}\>,
    \quad
    |1_L\> =\sum_{j\ {\rm odd}}
    c_j |D^{(k+1)n}_{(k+1)j}\>.
\end{align}
Here, the total number of qubits is $gn$, and we will strive to find the smallest number $n$ for which we can prove that the above PI code that corrects $k$ collective AD errors exists.
Evaluating the Knill-Laflamme conditions, the only inner products that end up being non-zero are of the form 
\begin{align}
\<D^{(k+1)n}_{(k+1)j}| ((J_-)^a)^\dagger
    (J_-)^a |D^{(k+1)n}_{(k+1)j}\>
\end{align}
for $a=0,\dots, k$.
 Therefore the A-matrix of \cite{ouyang2019permutation,movassagh2020constructing} will have $k+1$ rows and $n+1$ columns.
Any non-trivial solution to the homogeneous system of linear equations $A x = 0$ gives a solution for the PI code. For this, it suffices to have $n+1>k+1$, which is satisfied by setting $n= k+1$. Therefore, we can derive a  $(k+1)^2$-qubit PI code that corrects $k$ collective AD errors. 
For example, to correct one collective AD error, a 4-PI code is sufficient.

The PI-codes from this construction,
can be derived by first evaluating 
the expressions
\begin{align}
\<J,m|((J_-)^a)^\dagger (J_-)^a|J,m\>
= \Pi_{l=1}^{a} \gamma_{m-l},
\label{Jminus-power-inner-product}
\end{align}
 where \(\gamma_m = (J+m)(J-m+1)\),
 $J=(k+1)n/2$.
 Choosing $m= 
 -J + (k+1)j
 =
 (k+1)(j-n/2)$ for $j=0,\dots,n$, we calculate the $A$-matrix as
 \begin{align}
A = 
\sum_{a=0}^{k}
\sum_{j=0}^{n}
 \Pi_{l=1}^{a} \gamma_{(k+1)(j-n/2)-l+1} 
|a\>\<j|.
 \end{align}
For $k=1$, the A-matrix is 
 \begin{align}
A = 
\left(
\begin{array}{ccc}
 1 & 1 & 1 \\
 0 & 6 & 4 \\
\end{array}
\right)
 \end{align}
 with nullspace spanned by $(-1,-2,3)$.
Therefore, the logical codewords of the 4-qubit PI code which we call \textit{CAD4} (collective AD on four qubits) code that corrects one collective AD error  is:
 \begin{definition} \label{def:CAD4}
   CAD4 PI code: The logical codewords are
   \begin{equation}
       \begin{aligned}
           |0\rangle_L &:= |D^4_4\rangle, \\
|1\rangle_L &:= (|D^4_0\> + \sqrt 2|D^4_2\>)/\sqrt 3. 
       \end{aligned}
   \end{equation}
 \end{definition} 
For $k=2$, the A-matrix is 
 \begin{align}
     A = \left(
\begin{array}{cccc}
 1 & 1 & 1 & 1 \\
 0 & 21 & 24 & 9 \\
 0 & 336 & 600 & 144 \\
\end{array}
\right),
 \end{align}
 with nullspace spanned by 
 $(-4, -3, 0, 7)$. Therefore the logical codewords of the 9-qubit PI code or \textit{CAD9} code that corrects 2 collective AD errors has logical codewords: 
\begin{definition} \label{def:CAD9}
    CAD9 PI code: The logical codewords are 
    \begin{equation}
        \begin{aligned}
    |0\rangle_L &:= |D^9_9 \rangle, \\
|1\rangle_L &:= (
\sqrt {4/7} |D^9_0 \rangle
+
\sqrt{3/7} |D^9_3\>).
        \end{aligned}
    \end{equation}
\end{definition}
For any value of $k$, it is possible to solve for the nullspace of the corresponding A-matrix, though we do not have the analytical form of such a solution. The complexity of solving for the nullspace of the A-matrix is equal to the complexity of performing Gaussian elimination on a size $n$ matrix, which is $O(n^3)$.

\subsection{Implementing the recovery map using algorithm \ref{algo:cptp-implementation}} \label{sec: detailed implementation of the algorithm for AD noise}

We now apply the proposed algorithm \eqref{algo:cptp-implementation} 
to implement the recovery map $\mathcal{R}$ defined in 
Eq.~\eqref{eq: recovery kraus operators for AD channel}. Because the 
channel $\mathcal{R}$ has rank four, we require three ancilla qubits, 
which we denote as $a$, $b$, and $c$. Below, we follow the step by step, keeping track of the joint system--ancilla state after each branching and feedback operation.

\paragraph{Polar decomposition:}
\begin{align} \nonumber
R_r &= U_r B_r,
\ \ B_r := \sqrt{R_r^\dagger R_r} \ge 0,
\end{align}
for $r=0,1,2,3$. Here, \(B_r\) is a positive-semidefinite (hence Hermitian, \(B_r^\dagger = B_r\)) operator , and $U_r$ is a unitary operator. The motivation behind this decomposition is that, in the general, the Kraus operators are non-unitary and cannot be natively implemented on a closed system. Using this decomposition, the unitary part \(U_r\) is separated and can be implemented directly. Meanwhile, the non-unitary part \(B_r\), a positive, bounded operator satisfying \(B_r^2 \leq \mathbb{I}\) as mandated by the CPTP constraint, can be extended to larger unitary using ancilla and can be applied on the system as a controlled operation. 
\paragraph{Initialize}: \(
\sigma_{S,abc}
=
\rho_S \otimes |0\rangle_a\langle 0|
\otimes |0\rangle_b\langle 0|
\otimes |0\rangle_c\langle 0|
\).
\paragraph{Step 1) $\boldsymbol{r=0}$ \(\Rightarrow\) Binary split \(\{B_0,S_0\}\):}

Here, \(S_0 := \sqrt{\mathbb{I}-B_0^2}\). Then apply the unitary \(U_{S,a}^{(1)}\) on the system \(S\) and the first ancilla \(a\), given by
\begin{equation}
\label{eq: unitary_while_first_binary_splitting}
U_{S,a}^{(1)}
=
\begin{pmatrix}
B_0 & S_0 \\
S_0 & -B_0
\end{pmatrix}.
\end{equation}
The superscript \((1)\) denotes the step number. Now extend the unitary to full space as
\(
\mathbf{U}_{S,abc}^{(1)}
=
U_{S,a}^{(1)} \otimes \mathbb{I}_b \otimes \mathbb{I}_c \). The post-unitary state is
\begin{align}
\label{eq: state during first branch}
&\mathbf{U}_{S,abc}^{(1)} \sigma_{S,abc}
\bigl(\mathbf{U}_{S,abc}^{(1)}\bigr)^\dagger
\nonumber\\
&=
\Bigl[
B_0 \rho_S B_0 \otimes |0\rangle_a\langle 0|
+ B_0 \rho_S S_0 \otimes |0\rangle_a\langle 1| + 
\nonumber\\
&\qquad\;
S_0 \rho_S B_0 \otimes |1\rangle_a\langle 0|+ S_0 \rho_S S_0 \otimes |1\rangle_a\langle 1|
\Bigr] \\ \nonumber
& \qquad
\otimes |0\rangle_b\langle 0|
\otimes |0\rangle_c\langle 0|.
\end{align}
Now apply the feedback
\begin{align} \label{eq: feedback in first iteration}
C_{U_0}^{(a)}
&:=
\Bigl(
U_{0_S} \otimes |0\rangle_a\langle 0|
+
\mathbb{I}_S \otimes |1\rangle_a\langle 1|
\Bigr)
\otimes \mathbb{I}_b \otimes \mathbb{I}_c .
\end{align}
This gives
\begin{align}
\label{eq: state after first branching}
\sigma_{S,abc}^{(1)}
&=
\Bigl[
R_0 \rho R_0^\dagger \otimes |0\rangle_a\langle 0|
+ R_0 \rho S_0 \otimes |0\rangle_a\langle 1|
\nonumber\\
&\qquad\;
+ S_0 \rho R_0^\dagger \otimes |1\rangle_a\langle 0| \\  \nonumber
& + S_0 \rho S_0 \otimes |1\rangle_a\langle 1|
\Bigr]
\otimes |0\rangle_b\langle 0|
\otimes |0\rangle_c\langle 0|,
\end{align}
where we used \(U_0 B_0 = R_0\). \(\sigma_{S,abc}^{(1)}\) is the state after the first step which is denoted by the superscript \((1)\). 

\paragraph{Step 2) $\boldsymbol{r=1}$ \(\Rightarrow\) Binary split of the remainder into
\(\{\tilde{B}_1,\tilde{S}_1\}\)}:

Here, \(\tilde{B}_1 = B_1 S_0^{-1}\) and \(\tilde{S}_1 = S_1 S_0^{-1}\) with \(S_1 = \sqrt{B_2^2 + B_3^2}\). The updated Kraus operators follow completeness on the support of \(S_0\), i.e., \(
\tilde{B}_1^\dagger \tilde{B}_1 + \tilde{S}_1^\dagger \tilde{S}_1 = \mathbb{I}_{S_0}\). Now we apply an unitary 
\begin{align}
\label{eq: unitary while second binary splitting}
U_{S,b}^{(2)}
&=
\begin{pmatrix}
\tilde{B}_1 & * \\
\tilde{S}_1 & *
\end{pmatrix}, 
\end{align}
where the second column with \(*\) denotes an appropriate choice of orthogonal basis. The above unitary is extended to the full system such that it acts non-trivially on ancilla \(a\) in \(|1\rangle\) state (from now on we will call it \(a=1\) branch)  
\begin{equation}
\mathbf{U}_{S,abc}^{(2)}
=
|0\rangle_a\langle 0| \otimes \mathbb{I}_{S,bc}
+
|1\rangle_a\langle 1| \otimes U_{S,b}^{(2)} \otimes \mathbb{I}_c .
\end{equation}
Applying \(\mathbf{U}_{S,abc}^{(2)}\) to output state from the first iteration
\(\sigma_{S,abc}^{(1)}\) \eqref{eq: state after first branching}, and omitting terms that are
off-diagonal in the ancilla registers, gives
\begin{align}
\label{eq: state during second branch}
&\mathbf{U}_{S,abc}^{(2)}
\,\sigma_{S,abc}^{(1)}\,
\bigl(\mathbf{U}_{S,abc}^{(2)}\bigr)^\dagger
\nonumber\\
&=
R_0 \rho R_0^\dagger \otimes |000\rangle_{abc}\langle 000|
\nonumber\\
&\quad
+ B_1 \rho B_1 \otimes |100\rangle_{abc}\langle 100|
\nonumber\\
&\quad
+ S_1 \rho S_1 \otimes |110\rangle_{abc}\langle 110|
+ \text{off-diag.}
\end{align}
 On \(a=1\) branch, we used
\(\tilde{B}_1 S_0 = B_1\) and \(\tilde{S}_1 S_0 = S_1\).

Next apply the feedback on branch \(|100\rangle_{abc}\):
\begin{equation} \label{eq: feedback in second iteration}
C_{U_1} ^{(b)}
=
\mathbb{I}_S \otimes (\mathbb{I} - |100\rangle \langle 100|)_{abc}
+
U_1
\otimes |100\rangle_{abc}\langle 100|.
\end{equation}
The state becomes
\begin{align}
\label{eq: final state after second branch}
\sigma_{S,abc}^{(2)}
&=
R_0 \rho R_0^\dagger \otimes |000\rangle_{abc}\langle 000|
\nonumber\\
&\quad
+ R_1 \rho R_1^\dagger \otimes |100\rangle_{abc}\langle 100|
\nonumber\\
&\quad
+ S_1 \rho S_1 \otimes |110\rangle_{abc}\langle 110|
+ \text{off-diag.},
\end{align}
where we used \(U_1 B_1 = R_1\).

\paragraph{Step 3) \textbf{r=2} \(\Rightarrow\) Binary split of the final two Kraus operators \(\{\tilde{B}_2,\tilde{B}_3\}\)}, where \(\tilde{B}_2 = B_2 S_1^{-1} \) and \(\tilde{B}_3 = B_3 S_1^{-1}\) following the completeness condition on the support of \(S_1\). 
Now we apply a unitary \begin{align}
\label{eq: unitary while third binary splitting}
U_{S,c}^{(3)}
&=
\begin{pmatrix}
\tilde{B}_2 & * \\
\tilde{B}_3 & *
\end{pmatrix},
\end{align} 
where as mentioned above \(*\) denotes a choice of appropriate orthogonal basis. Now this unitary is extended to the full system acting non-trivially only on the branch \(a=1\), \(b=1\), i.e.
\begin{equation}
\mathbf{U}_{S,abc}^{(3)}
=
\mathbb{I}_{S,abc}
+
|11\rangle_{ab}\langle 11|
\otimes \bigl(U_{S,c}^{(3)}-\mathbb{I}_{S,c}\bigr).
\end{equation}
Applying it to the output state from last step \(\sigma_{S,abc}^{(2)}\) \eqref{eq: final state after second branch} gives
\begin{align}
\label{eq:state_during_third_branch}
&\mathbf{U}_{S,abc}^{(3)}
\,\sigma_{S,abc}^{(2)}
\,\bigl(\mathbf{U}_{S,abc}^{(3)}\bigr)^\dagger
\nonumber\\
&=
R_0 \rho R_0^\dagger
\otimes |000\rangle_{abc}\langle 000|
\nonumber\\
&\quad
+ R_1 \rho R_1^\dagger
\otimes |100\rangle_{abc}\langle 100|
\nonumber\\
&\quad
+ B_2 \rho B_2
\otimes |110\rangle_{abc}\langle 110|
\nonumber\\
&\quad
+ B_3 \rho B_3
\otimes |111\rangle_{abc}\langle 111|
+ \text{off-diag.}
\end{align}

Applying the feedback one final time on remaining branches \(b=1\) and \(c=1\), we get
\begin{align} \label{eq: feedback in final iteration}
C_{U_2,U_3} ^{(b),(c)}
&=
\mathbb{I}_{S,abc}
+
\bigl(U_2-\mathbb{I}_S\bigr)
\otimes |110\rangle_{abc}\langle 110|
\nonumber\\
&\quad
+
\bigl(U_3-\mathbb{I}_S\bigr)
\otimes |111\rangle_{abc}\langle 111|.
\end{align}
Note that we used the relation \(
\mathbb{I} \otimes \mathbb{I} + \bigl(U-\mathbb{I}\bigr)\otimes P =
\mathbb{I}\otimes\bigl(\mathbb{I}-P\bigr)+U\otimes P
\) to compactly write the unitaries and controlled-unitaries above, where \(P\) is a projector.  
The final state is
\begin{align}
\rho_{\mathrm{final}}
&=
R_0 \rho R_0^\dagger \otimes |000\rangle_{abc}\langle 000|
\nonumber\\
&\quad
+ R_1 \rho R_1^\dagger \otimes |100\rangle_{abc}\langle 100|
\nonumber\\
&\quad
+ R_2 \rho R_2^\dagger \otimes |110\rangle_{abc}\langle 110|
\nonumber\\
&\quad
+ R_3 \rho R_3^\dagger \otimes |111\rangle_{abc}\langle 111|,
\end{align}
where we used \(U_2 B_2 = R_2\) and \(U_3 B_3 = R_3\).

Tracing out the ancillas \(abc\) yields the desired CPTP map:
\begin{align}
\mathcal{R}(\rho)
&=
\mathrm{Tr}_{abc}\!\left[\rho_{\mathrm{final}}\right]
\nonumber\\
&=
R_0 \rho R_0^\dagger
+ R_1 \rho R_1^\dagger
+ R_2 \rho R_2^\dagger
+ R_3 \rho R_3^\dagger .
\end{align}

\section{Hardware efficient implementation using Geometric Phase gates}
\subsection{Compiling unitaries} \label{sec: compiling unitaries}

We now provide an explicit gate-level decomposition of the unitaries in Eqs.~\eqref{eq: unitary_while_first_binary_splitting}, \eqref{eq: unitary while second binary splitting}, and \eqref{eq: unitary while third binary splitting}, together with the controlled unitaries in Eqs.~\eqref{eq: feedback in first iteration}, \eqref{eq: feedback in second iteration}, and \eqref{eq: feedback in final iteration} into native operations available through geometric phase gates (GPGs) to implement the algorithm. In the context of permutation-invariant (PI) codes, the system \(S\) is restricted to the \((N+1)\)-dimensional Dicke subspace, where \(N\) is the number of physical qubits, and each ancilla is a single qubit. The following operations can be implemented efficiently:
\begin{enumerate}
    \item exact arbitrary unitaries (and state synthesis)~\cite{jandura2024nonlocal, johnsson2020geometric, srivastava2026entanglement} and approximate unitaries \cite{gutman2024universal} on the Dicke subspace by coupling the qubits to a bosonic mode;
    \item arbitrary single-qubit gates; and
    \item entangling controlled system-ancilla unitary of the form 
    \begin{equation} 
        \Lambda(U)_{S,A}
        :=
        I_S \otimes |0\rangle_A\langle 0|
        +
        U \otimes |1\rangle_A\langle 1|.
    \end{equation} 
\end{enumerate}

In our setup, \(A\) is a single-qubit ancilla. During compilation, we use the following standard single-qubit gates
\begin{equation} \nonumber
    H = \frac{1}{\sqrt{2}}
    \begin{pmatrix}
        1 & 1\\
        1 & -1
    \end{pmatrix},
    \qquad
    P =
    \begin{pmatrix}
        1 & 0\\
        0 & i
    \end{pmatrix},
    \qquad
    Z =
    \begin{pmatrix}
        1 & 0\\
        0 & -1
    \end{pmatrix}.
\end{equation}

The unitary in Eq.~\eqref{eq: unitary_while_first_binary_splitting} can be decomposed as 
\begin{align} \label{eq:decomposing_first_unitary}
U_{S,a}^{(1)}(\Theta_1)
={}&
\left(\left(e^{i\Theta_1}\right)_S \otimes I_a\right)
\left(I_S \otimes P_a^\dagger\right)
\left(I_S \otimes H_a\right)
\nonumber\\
&\times
\Lambda\!\left(e^{-2i\Theta_1}\right)_{S,a}
\left(I_S \otimes H_a\right)
\left(I_S \otimes P_a Z_a\right).
\end{align}

where $\cos\Theta_1=B_0$ and $\sin\Theta_1=S_0$ while \(\Lambda(e^{-2i \Theta_1})_{S,a} = \mathbb{I}_S\otimes |0\rangle_a \langle 0| + e^{-2i \Theta_1}_S \otimes |1\rangle_a \langle1|\). Note that if ancilla starts in the \(\ket{0}_a\) state, we can drop the first \(Z_a\) gate. \newline
Now we will use a slight modification of the above method to decompose the second unitary given in Eq.~\eqref{eq: unitary while second binary splitting}. Since $\tilde{B}_1$ and $\tilde{S}_1$ are both Hermitian operators and commute ($[\tilde{B}_1, \tilde{S}_1]=0$), we can apply spectral decomposition to write $\tilde{B}_1^\dagger\tilde{B}_1 = F C^2 F^\dagger$ and $\tilde{S}_1^\dagger\tilde{S}_1 = F D^2 F^\dagger$, where $F$ is a unitary operator. Clearly, $C^2 + D^2 = I$ due to the completeness condition. From the spectral decomposition, the positive diagonal matrices are given by $C = F^\dagger \tilde{B}_1 F$ and $D = F^\dagger \tilde{S}_1 F$. We can then define the operators $L_{10} := \tilde{B}_1 F C^{-1}$ and $L_{11} := \tilde{S}_1 F D^{-1}$, which are extended to unitaries acting on the full Dicke space. Concretely, we can write the first column of the \(U_{S,b}^{(2)}\) \eqref{eq: unitary while second binary splitting} as 
\begin{equation}
    \begin{pmatrix}
    \tilde{B}_1 \\ 
    \tilde{S}_1
    \end{pmatrix} = \begin{pmatrix}
        L_{10} & 0 \\  
        0 & L_{11} 
    \end{pmatrix} \begin{pmatrix}
        C \\ 
        D 
    \end{pmatrix} F^{\dagger}.
\end{equation} 
We decompose the left hand side as
\begin{align} \label{eq: decomposing second unitary}
   U_{S,b}^{(2)}(\Theta_2)= (L_{{10}_S} \otimes I_b ) \Lambda (L_{10}^{\dagger} L_{11})_{S,b} U_{S,b} ^{(1)} (\Theta_2) (F^{\dagger}_S \otimes I_b)
\end{align}
where \(\cos(\Theta_2) = C \), \(\sin(\Theta_2) = D\) and  \(\Lambda (L_{10}^{\dagger} L_{11})_{S,b}= \mathbb{I}_S \otimes |0\rangle _b \langle 0| + (L_{10}^{\dagger} L_{11})_S \otimes |1\rangle_b\langle 1|\) is the ancilla controlled operation on the system. We already know the decomposition of \(U_{S,a}^{(1)}(\Theta_2)\) \eqref{eq:decomposing_first_unitary}. For the third unitary in Eq.~\eqref{eq: unitary while third binary splitting} we repeat the same process as we did here. \newline

First iteration \(r=0\) requires a single entangling gate
$\Lambda(e^{-2i\Theta_1})_{S,a}$ while, second \(r=1\) and third \(r=2\) iteration each require two entangling gates: $\Lambda(e^{-2i\Theta_j})_{S,b}$ and $\Lambda(L_{j0}^\dagger L_{j1})_{S,c}$ for $j\in\{2,3\}$. All other operations are either system-only unitaries on the Dicke subspace or single-qubit ancilla gates.

\begin{figure*}[t]
\centering
\includegraphics[width=0.95\textwidth]{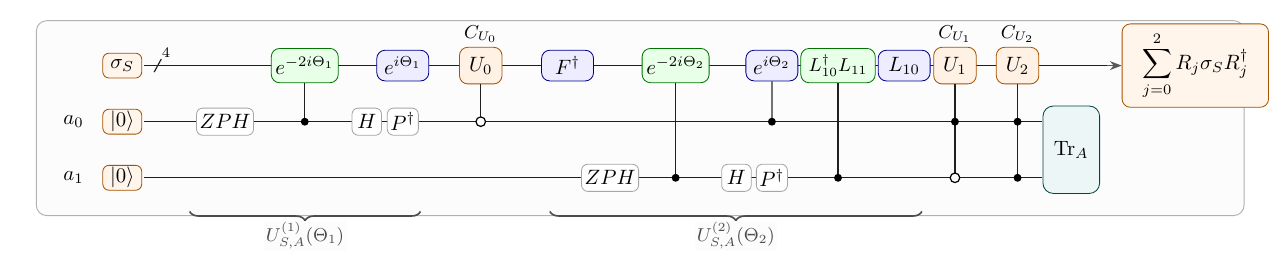}
\caption{
Compiled recovery circuit for CAD4 PI code given in Definition~\ref{def:CAD4}, correcting a first-order global symmetric correlated AD error. The circuit closely follows Algorithm~\ref{algo:cptp-implementation}. The compilation of the first system--ancilla unitary \(U^{(1)}_{S,A}(\Theta_1)\) is given in Eq.~\eqref{eq:decomposing_first_unitary}, while the compilation of the second system--ancilla unitary \(U^{(2)}_{S,A}(\Theta_2)\) is given in Eq.~\eqref{eq: decomposing second unitary}. The parameters \(\Theta_1\) and \(\Theta_2\) are determined by the noise strength.}
\label{fig:cad4-recovery-circuit}
\end{figure*}

As a small example of the compilation scheme, Fig.~\ref{fig:cad4-recovery-circuit} shows an explicit recovery circuit for CAD4 PI code to recover from global symmetric AD noise up to first order. 

\subsection{Geometric-phase-gate implementation of compiled unitaries}
\label{sec:gpg-implementation-elementary-ops}

We need to show preparation of an arbitrary unitary \(U \in \mathrm{SU(N+1)}\) and a controlled system-ancilla unitary \(\Lambda(U)\), as mentioned above. We saw in the previous section that if we are able to do these primitives then we can implement all the required unitaries. We can write an arbitrary unitary in Dicke subspace as \(U  = \sum_{l=0} ^{N} e^{i\phi_l} P_l\)
where \(P_l=|u_l\rangle \langle u_l|\) and \(\{|u_l\rangle\}_{l=0} ^{N}\) are orthogonal basis set spanning the Dicke subspace. Now, \([P_l, P_m]=0\)  \(\forall l, m\), we can rewrite \(U\) as 
\begin{align} \label{eq: a general unitary}
    U & = \prod_{l=0}^{N} e^{i \phi_l P_l}. \\ \nonumber 
\end{align}
We can rewrite \(P_l\) in terms of unitary extension synthesized via GPGs as 
\begin{align}
    P_l = W^{\dagger}_l|D^N_N \rangle \langle D^N_N| W_l, 
\end{align}
using \(|u_l\rangle = W_l^{\dagger} |D^N_N\rangle\) where \(W_l^{\dagger}\) is a unitary for preparing state \(|u_l\rangle\) and hence called state-preparation unitary. 
Upon substituting in Eq.~\eqref{eq: a general unitary} we get 
\begin{align} \label{eq:form of a general unitary in terms of GPG pulses}
    U & = \prod_{l=0}^{N} e^{i \phi_l W^{\dagger}_l|D^N_N \rangle \langle D^N_N| W_l} \\ \nonumber 
    & = \prod_{l=0}^{N} W^{\dagger}_l e^{i \phi_l |D^N_N \rangle \langle D^N_N|} W_l \\\nonumber 
    & = \prod_{l=0}^{N} W^{\dagger}_l \Phi_N (l) W_l, 
\end{align}
where we used \(\Phi_N (l) = e^{i \phi_l |D^N_N \rangle \langle D^N_N|}\) for simplicity.\newline
Now we need to show similar implementation of \(\Lambda(U)\) which is 
\begin{align} \label{eq: form of a general controlled unitary in terms of GPG pulses}
    \Lambda(U) & = \mathbb{I} \otimes |0\rangle \langle0| + U \otimes |1\rangle \langle 1| \\ \nonumber 
    & = \mathbb{I} \otimes |0\rangle \langle 0| + \prod_{l=0}^{N} W^{\dagger}_l e^{i \phi_l |D^N_N \rangle \langle D^N_N|} W_l \otimes |1\rangle \langle 1| \\ \nonumber 
    & = \prod_{l=0}^{N} W^{\dagger}_l e^{i \phi_l |D^{N+1}_{N+1}\rangle \langle D^{N+1}_{N+1}|} W_l \\ \nonumber 
    & = \prod_{l=0}^{N} W^{\dagger}_l \Phi_{N+1} (l) W_l,  
\end{align}
where \(\Phi_{N+1} (l) = e^{i \phi_l |D^{N+1}_{N+1} \rangle \langle D^{N+1}_{N+1} |} \). 

\begin{table}[htbp] 
    \centering
    \renewcommand{\arraystretch}{1} 
    \begin{tabular}{|c|c|}
        \hline
        \textbf{Gate} & \textbf{Time complexity}  \\ 
        \hline
        $W_l$\ \text{state\    preparation} & $2N$ \\ 
        \hline
        $e^{i \phi_l |D^N_N\rangle \langle D^N_N|}$ \text{multi-controlled phase} & $N-1$  \\ 
        \hline
        \text{permutation symmetric} $U\in \mathrm{SU(N+1)}$ & \(5N^2-N\)  \\ \hline 
    \end{tabular}
    \caption{Time steps needed for preparation of elementary operations we require while unitary implementation. \(N\) denotes the number of spins. Note that in the given time steps we get an exact gate.}
     \label{tab: time complexity of elementary operations}
\end{table}

Table.~\ref{tab: time complexity of elementary operations} summarizes the gate complexity of the elementary operations. In particular, the state-synthesis unitary \(W_l\), as well as its inverse \(W_l^\dagger\), can each be implemented exactly in \(2N\) GPG steps \cite{ouyang2025measurement, Bond2025}. Likewise, the special unitary \(\Phi_{N+1}(l)\) can be implemented exactly in \(N-1\) GPG steps \cite{ouyang2025measurement}. Since an arbitrary unitary of the form given in Eq.~\eqref{eq:form of a general unitary in terms of GPG pulses} is constructed from the composition of \(W_l\), \(W_l^\dagger\), and \(\Phi_{N+1}(l)\), its total GPG complexity is
\begin{align}
n_{\mathrm{GPG}}(U)
&= N\Bigl[
n_{\mathrm{GPG}}(W_l)
+ n_{\mathrm{GPG}}(W_l^{\dagger})
+ n_{\mathrm{GPG}}\bigl(\Phi_{N+1}(l)\bigr)
\Bigr]
\nonumber\\
&= N\bigl[2N + 2N + (N-1)\bigr]
\nonumber\\
&= 5N^2 - N .
\label{eq:time_complexity_for_arbitrary_U}
\end{align}

\subsection{State preparation}

An arbitrary target state $|\Psi_T\rangle$ within the Dicke subspace can be prepared from an initial state, here we take all-one state $|D^N_N\rangle \equiv |J=\frac{N}{2}, M=-\frac{N}{2}\rangle$, via a sequence of GPGs and collective rotations. The ideal state preparation is given by
\begin{align} \label{eq: state preparation using GPGs}
    \ket{\Psi} = \prod_{p=1}^{P} \Bigg[ R(\alpha_p,\beta_p,\gamma_p) \mathrm{U}_{\mathrm{GPG}}(\phi_p) \Bigg] |D^N_N\rangle, 
\end{align}
where the collective spatial rotation is defined as
\begin{align} \nonumber
    R(\alpha_p,\beta_p,\gamma_p) = R_z(\alpha_p)R_y(\beta_p)R_x(\gamma_p).
\end{align}
Here, $R_n(\theta) = e^{-i \theta J_n}$ for $n \in \{x,y,z\}$, $J_n$ are collective spin operators, and $\theta$ denotes the rotation angle. $P$ is the total number of pulse sequences required, and $\mathrm{U}_{\mathrm{GPG}}$ represents the geometric phase gate, which can be linear (as in Ref.~\cite{molmer1999multiparticle}) or non-linear (as in Ref.~\cite{johnsson2020geometric}). Each pulse sequence $p$ consists of a GPG followed by global rotations. The exact pulse parameters $(\alpha_p, \beta_p, \gamma_p, \phi_p)$ are determined numerically via quantum optimal control to minimize the state infidelity $1 - |\langle\Psi | \Psi_T\rangle|^2$. Successive pulse sequences steer the system toward the target state. To construct the arbitrary unitaries and controlled-feedback operations required for our protocol, we perform a spectral decomposition of the target operators and utilize this GPG sequence to independently prepare the eigenvectors corresponding to non-zero eigenvalues, as detailed in Eqs.~\eqref{eq:form of a general unitary in terms of GPG pulses} and~\eqref{eq: form of a general controlled unitary in terms of GPG pulses}.

GPGs are realized in a laboratory by coupling a register of spins to a coherently controllable bosonic mode, such as a cavity field or a motional mode in an ion trap. This mode acts as a mediator for entanglement, and its frequency, driving intensity, and detuning can be dynamically controlled. If the spin-mode coupling is linear in the cavity mode's creation and annihilation operators, the resulting operation is termed a linear GPG. A prominent example is the M\o lmer-S\o rensen-type gate \cite{molmer1999multiparticle}, which generates an effective unitary of the form $e^{i \phi J_z^2}$. When complemented with global spin rotations, this interaction enables efficient, exactly universal state and unitary synthesis within the Dicke subspace \cite{gutman2024universal, Bond2025, srivastava2026entanglement}. Other linear GPGs have also been developed \cite{jandura2024nonlocal}, offering a broader class of entangling operations, such as the multi-controlled phase gate ($\mathrm{C_{N-1}(Z)}$). By contrast, GPGs that rely on dispersive coupling, which is quadratic in the bosonic mode operators, are classified as non-linear GPGs. When paired with collective rotations, these gates also form an exactly universal set for synthesis in the Dicke subspace \cite{johnsson2020geometric}. 

In this work, we utilize a linear GPG of the type $e^{i\phi J_z^2}$. In practical implementations, however, the target spins can spontaneously decay and the mediating bosonic mode is susceptible to leakage (for detailed derivation of the open-system Hamiltonian governing these dynamics, see Ref.~\cite{jandura2024nonlocal}). Consequently, the GPG operations become noisy. To accurately model this decoherence and the associated loss of probability from the logical subspace, the ideal $p$-th unitary $\mathrm{U}_{\mathrm{GPG}}(\phi_p)$ acting on the state $\rho_{p-1}$ is replaced by a completely positive, non-trace-preserving map $\mathcal{E}_{\mathrm{GPG}}(\phi_p)[\rho_{p-1}]$, defined as
\begin{align}
\mathcal{E}_{\mathrm{GPG}}(\phi_p)[\rho_{p-1}] &= \sum_{n,m=0}^{N} \bra{D_n^N}\rho_{p-1}\ket{D_m^N} \nonumber \\
&\quad \times e^{i\Phi_{n,m}(\phi_p)} \ket{D_n^N}\bra{D_m^N},
\end{align}
where the phase function $\Phi_{n,m}(\phi_p)$ is given by
\begin{equation}
\Phi_{n,m}(\phi_p) = \Bigg[ n^2-m^2 + (n-m)^2\frac{i \kappa}{2\delta} + (n+m) \frac{i\gamma \delta}{2g^2} \Bigg] \phi_p.
\label{eq:gpg-fnm}
\end{equation}
Here, $\kappa$ denotes the decay rate of the cavity mode, $\gamma$ is the spontaneous decay rate of the atoms, $g$ is the atom-cavity coupling strength, and $\phi_p$ corresponds to the phase acquired by an ideal GPG in the absence of losses (i.e., $\gamma = \kappa = 0$). A relevant parameter is cavity cooperativity $C = g^2/(\kappa \gamma)$, which parametrizes the ratio of the coherent coupling strength to the decay rates. Defining the ratio $a = \gamma/\kappa$, we get $g^2 = C\kappa^2 a$. By setting $a=1$ and working in units of energy where $g=1$, we find $\kappa = \gamma = 1/\sqrt{C}$. Substituting these values into Eq.~\eqref{eq:gpg-fnm} yields
\begin{equation}\label{eq:noisy_gpg_equation_final} 
\Phi_{n,m}(\phi_p) = \Bigg[ (n^2-m^2) + \frac{i(n-m)^2}{2\sqrt{C} \delta} + \frac{i(n+m)\delta}{2\sqrt{C}} \Bigg] \phi_p .
\end{equation}

Applying the subsequent collective rotations yields the complete noisy map for the $p$-th pulse sequence:
\begin{equation}
\mathcal{E}^{(p)}[\rho_{p-1}] = R(\alpha_p,\beta_p,\gamma_p) \mathcal{E}_{\mathrm{GPG}}(\phi_p)[\rho_{p-1}] R^{\dagger}(\alpha_p,\beta_p,\gamma_p).
\end{equation}
Note that we assume the global collective rotations are error-free, as single-qubit and global microwave or optical pulses can typically be driven with fidelities orders of magnitude higher than the slower, mode-mediated GPG entangling operations. Finally, the full state preparation under experimental noise is governed by the sequential composition of these maps: \(\mathcal{E}_{\mathrm{total}} = \mathcal{E}^{(P)} \circ \mathcal{E}^{(P-1)} \circ \cdots \circ \mathcal{E}^{(1)}\). 

\section{Numerical Results}
\label{sec:numerical_results for error recovery}

\subsection{Implementation of Algorithm 1} \label{sec:implementation of algorithm 1}
\begin{figure}[h!]
    \centering
    \includegraphics[width=\linewidth]{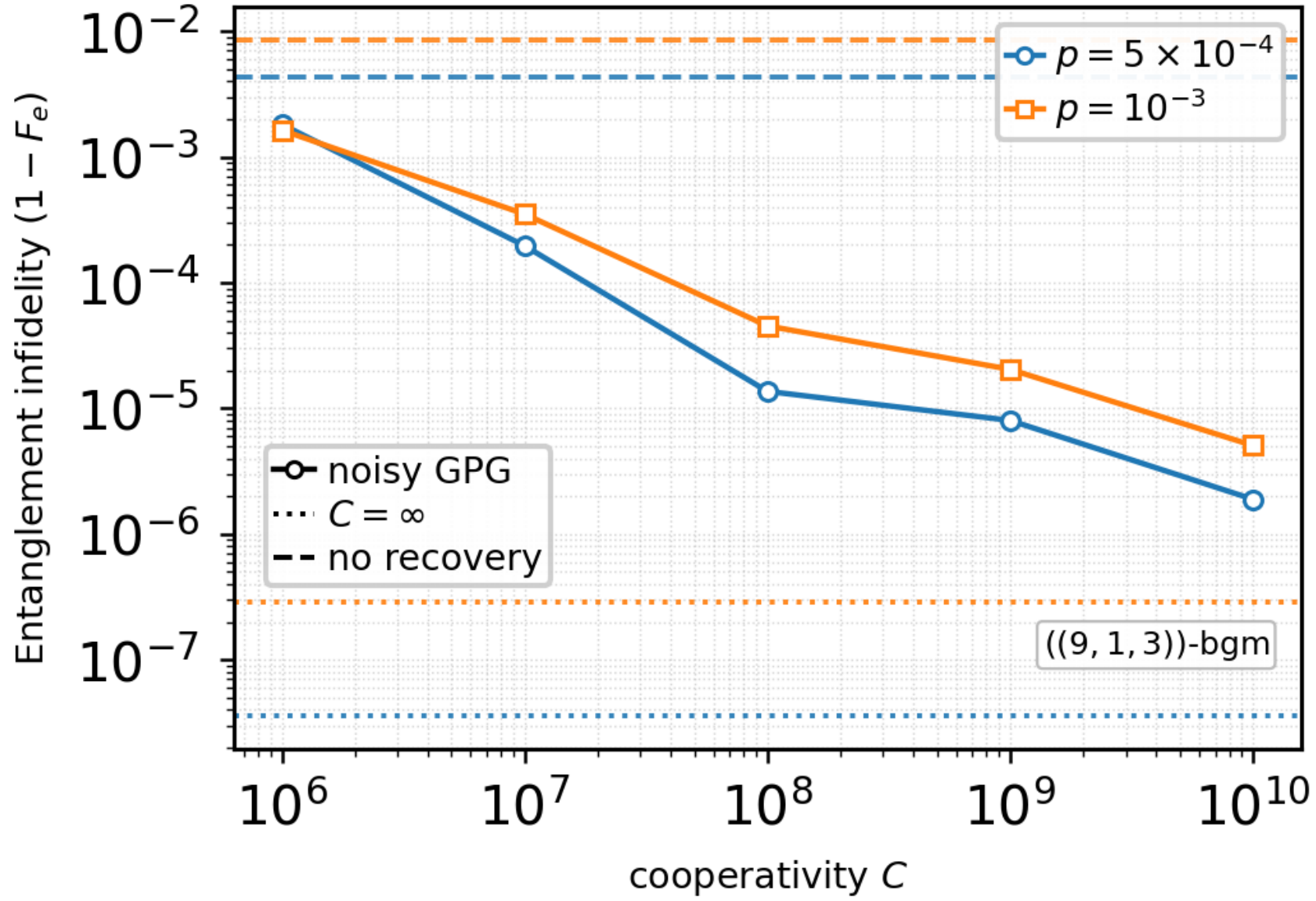} 
    \caption{
Entanglement infidelity vs cooperativity after implementing the recovery Algorithm~\ref{algo:cptp-implementation} on
\(((9,1,3))\)-bgm code under exact global symmetric correlated AD noise. The recovery map in Eq.~\eqref{eq: recovery kraus operators for AD channel} is computed using the Petz recovery formula corresponding to the second-order approximation of the noise channel in Eq.~\eqref{app:final set of kraus operators for truncated dynamics}. The recovery map is then implemented using Algorithm \ref{algo:cptp-implementation}, with the compilation scheme described in Sec.~\ref{sec: compiling unitaries} and the noisy linear GPG sequence in Eq.~\eqref{eq:noisy_gpg_equation_final}. Solid curves show the GPG-based recovery implementation, while the dotted curves at the bottom show the recovery performance in the absence of losses \(\gamma =\kappa =0\) and \(C \rightarrow \infty \). The dashed curve at the top represents the performance of a bare qubit without any recovery. The comparison is shown for two experimentally relevant AD noise
strengths, \(p=5.0\times 10^{-4}\) and
\(p=5.0\times 10^{-3}\).}
    \label{fig:final recovery using algorithm} 
\end{figure}

We consider the explicit example of the \(((9,1,3))\)-\(\mathrm{bgm}\) code. After the encoded state undergoes exact global symmetric AD noise, we numerically implement the recovery map in Eq.~\eqref{eq: recovery kraus operators for AD channel}, obtained from the second-order truncated dynamics, using Algorithm~\ref{algo:cptp-implementation}. We use the noisy GPG sequence \eqref{eq:noisy_gpg_equation_final} and the result is as shown in Fig.~\ref{fig:final recovery using algorithm}.

We find that the recovery scheme gives several orders of suppression in entanglement infidelity compared with doing no recovery, even when the gates are noisy. The suppression improves as the cooperativity \(C\) increases, corresponding to better-quality gates with reduced loss. For reference, the dotted curve in the bottom panel shows the ideal recovery limit, where \(\gamma=\kappa=0\) and \(C\rightarrow\infty\). At first sight, the cooperativities used here may seem high for current experimental platforms. However, we stress that collective cooperativities in the range \(C\sim 10^3\)--\(10^4\) have already been demonstrated in optical cavities.~\cite{sauer2004cavity,tuchman2006normal,kawasaki2019geometrically}. Therefore, the parameter regime considered here is demanding, but not unrealistic, and continued improvements in optical-cavity platforms could bring these values within reach for high-fidelity implementations.

For details of the implementation, see the Github repository \cite{chandra_qer_pi_2026}.  

\subsection{Error recovery} \label{sec:numerical results for optimum error recovery}
In this section, we numerically evaluate the performance of the PI codes listed in Table~\ref{tab:pi-codes-studied} along with our newly discovered CAD PI codes under both global and local symmetric AD noise. As mentioned before, we use entanglement fidelity as the metric to gauge the performance of the recovery operation by varying the error probability \(p = \gamma dt\). Simulating local symmetric noise typically requires tracking the full \(2^N\)-dimensional Hilbert space, since the independent single-qubit Pauli errors can cause the state to leak out of the Dicke subspace. However, because these local errors act identically on all physical qubits, we can exploit the Schur-Weyl decomposition to represent their action in the collective basis. This symmetry-adapted representation exponentially reduces the dimensionality of the simulation from \(2^N\) to just \(O(N^2)\) \cite{Shammah2018PIQS}. For a detailed discussion of this mapping, see Appendix~\ref{app: collective basis representation}. \newline 

\begin{figure}
    \centering
    \includegraphics[width=\linewidth]{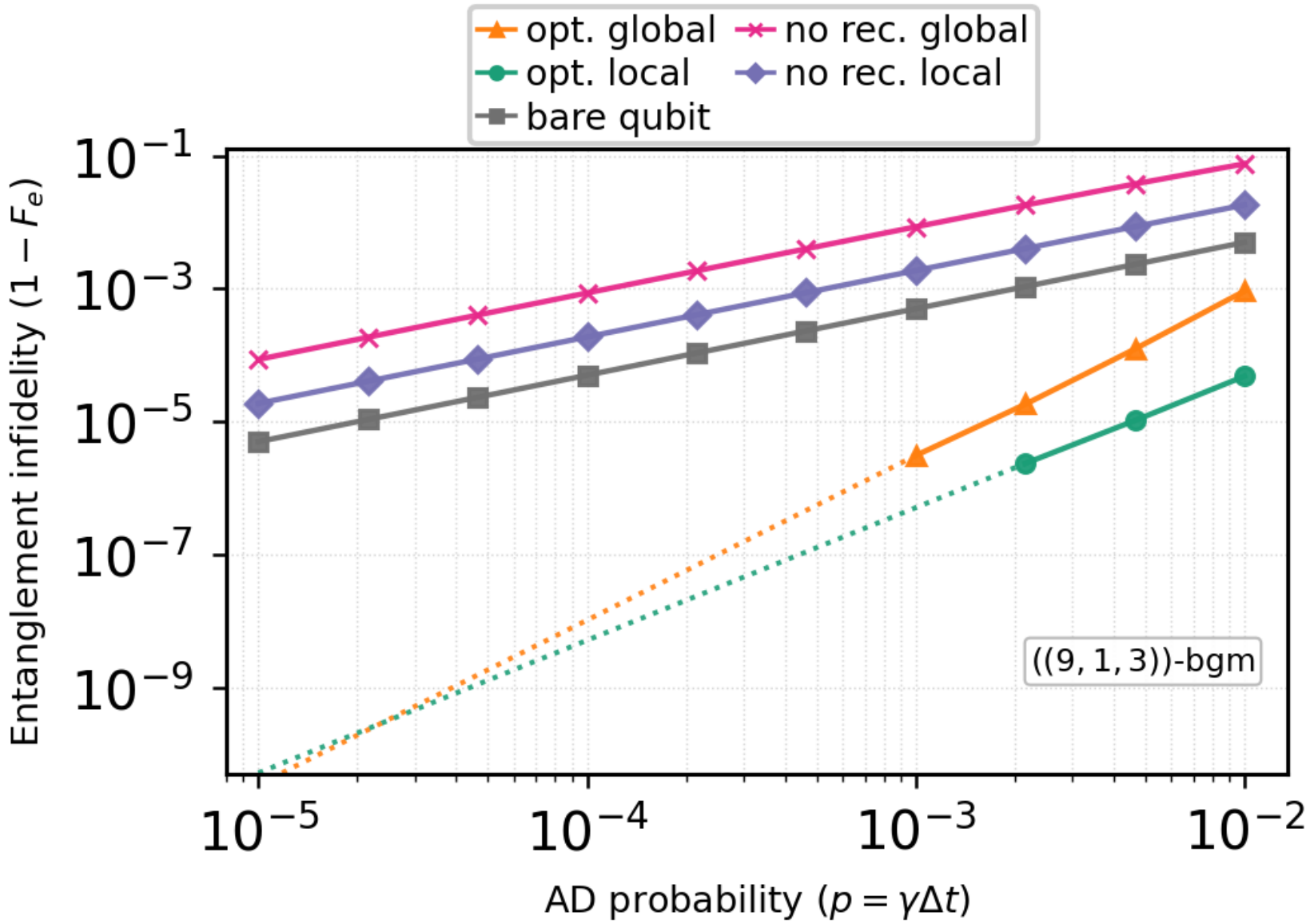}
    \caption{Performance of the $((9,1,3))$ bgm code ($b=g=3, m=1$) under global and local symmetric AD noise channels. Optimal SDP-based recovery (``opt.'') is contrasted against uncorrected states (``no rec.'') and an unencoded bare qubit baseline. Active error recovery yields orders-of-magnitude suppression in infidelity, whereas omitting recovery renders the encoded state strictly worse than the bare qubit. Note that the global symmetric correlated noise consistently causes more harm than local symmetric iid noise.}
    \label{fig:bgm_code_global_local_AD}
\end{figure}

To the best of our knowledge, this work represents the first explicit numerical benchmarking of QER for PI codes. We start our numerical analysis by demonstrating the benefit of optimized error recovery and showing the different behavior of PI codes under global and local symmetric AD noise. Fig.~\ref{fig:bgm_code_global_local_AD} illustrates the performance of the $((9,1,3))$ bgm code, highlighting two critical phenomena that frame our broader analysis. First, applying an optimal recovery map suppresses the infidelity by orders of magnitude. Conversely, without active correction, the encoded state performs strictly worse than an unencoded bare qubit, confirming that encoding into a larger physical Hilbert space is inherently detrimental unless paired with an effective recovery channel. Second, we observe that global symmetric noise induces more rapid decoherence than local symmetric noise. Physically, the global channel implies the environment cannot distinguish between individual spins, treating the system as a single collective entity. This interaction is governed by collective operators $J_k$, whose matrix elements scale up to $\mathcal{O}(N)$ near the middle of the Dicke ladder. Consequently, the effective noise strength, proportional to the square of these elements, scales quadratically as $\mathcal{O}(N^2)$, a phenomenon related to superradiance. In contrast, local symmetric noise acts on individual spins. Because the underlying error mechanism is strictly single-particle in nature, it acts via standard Pauli operators with $\mathcal{O}(1)$ matrix elements. Even when summing these independent errors over all $N$ qubits, the total error rate scales only linearly as $\mathcal{O}(N)$. Therefore, although local noise forces the state to leak out of the Dicke subspace (unlike global noise, which preserves it), the global noise causes a significantly faster loss of fidelity due to this $\mathcal{O}(N^2)$ collective enhancement factor.

\begin{figure}
    \centering
    \includegraphics[width=\linewidth]{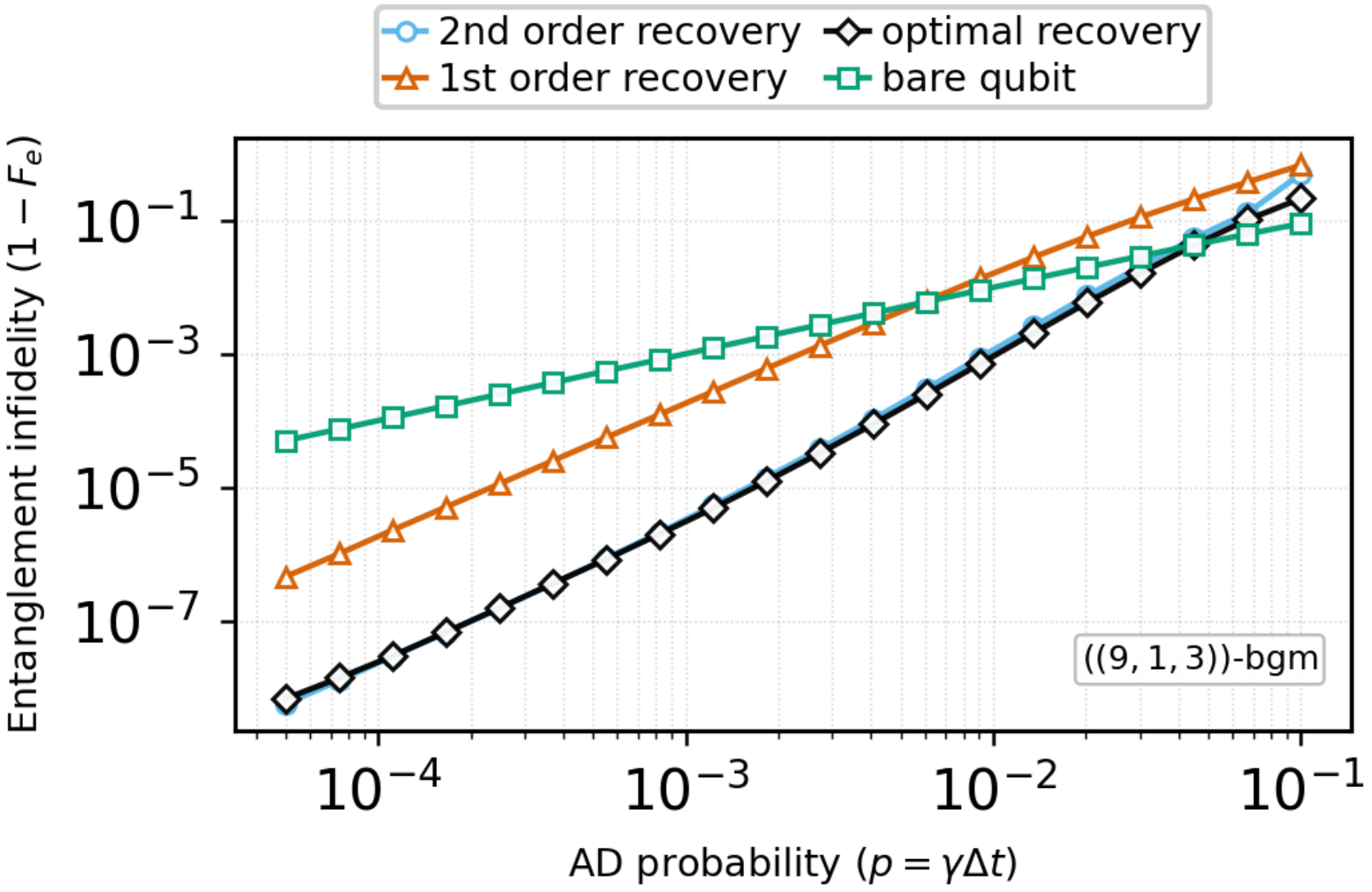}
    \caption{Effectiveness of QER against correlated AD errors. The state encoded in codespace of \(((9,1,3)) \mathrm{bgm}\) code is evolved under the exact global symmetric correlated AD channel. Then we compare three recovery maps: the SDP-optimal recovery for the full channel (black), the second-order Petz recovery (light blue), and the first-order Petz recovery (red). A bare qubit going through the same channel with Petz recovery is also shown for comparison (green). The first-order recovery is several orders of magnitude worse than the exact SDP recovery, while the second-order recovery almost overlaps with it. This shows that higher order correlated AD errors give dominant contribution in this regime and are successfully corrected by QER.}
    \label{fig:penalty}
\end{figure}

\begin{figure*}[t]
    \centering
    \includegraphics[width=\linewidth]{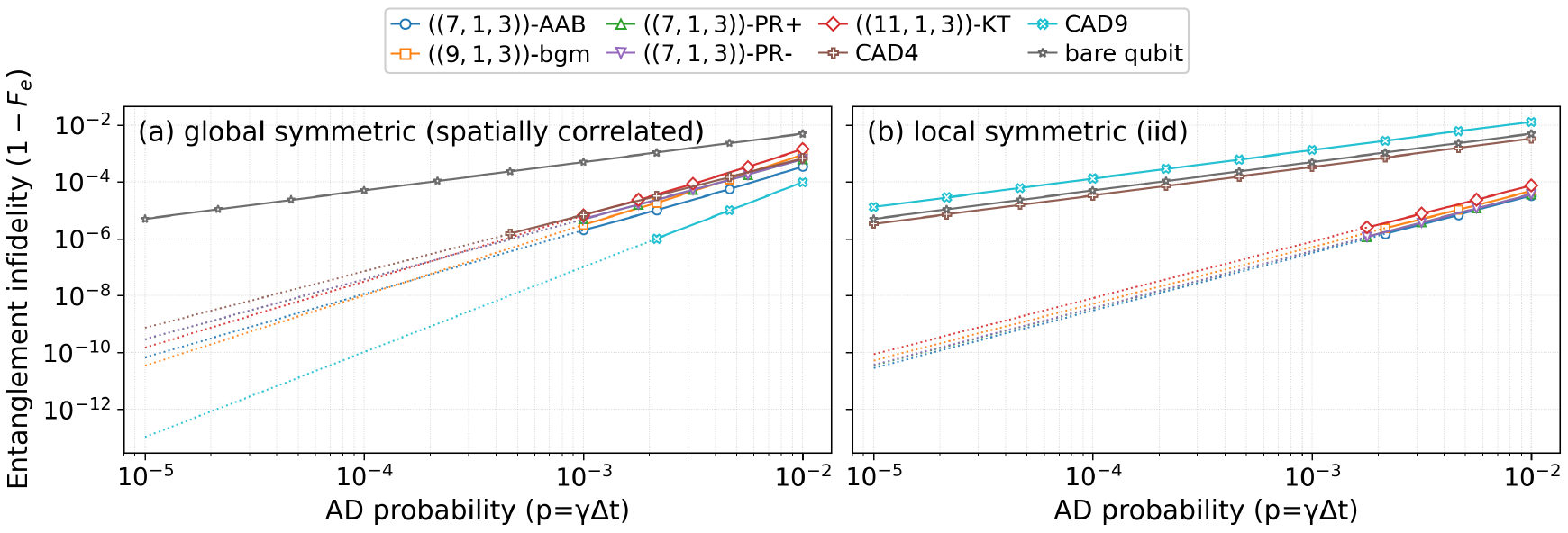}
    \caption{ Performance of the short-length PI codes listed in Table~\ref{tab:pi-codes-studied}, together with the newly introduced CAD4 and CAD9 PI codes, under optimal SDP quantum error recovery. (a) The left panel shows the case of global symmetric correlated AD noise. In this setting, the CAD9 PI code gives the best performance and suppresses the entanglement infidelity by several orders of magnitude compared with the second-best code, the 7-qubit AAB code. (b) The right panel shows the case of local symmetric, or iid, AD noise. Here the CAD PI codes perform badly, since they are not optimized for iid AD noise. Instead, the 7-qubit AAB code gives the strongest performance, consistent with the usual distance-based condition when the errors are independent and localized. Solid lines denote calculated data points, while dotted lines represent extrapolations.}
    \label{fig:all_codes_benchmark_global_local}
\end{figure*}

Fig.~\ref{fig:all_codes_benchmark_global_local} compares the performance of the already existing short-length PI codes listed in Table~\ref{tab:pi-codes-studied} with the newly discovered CAD codes under local and global symmetric AD noise. The left panel shows the case of global symmetric correlated AD case while the right one shows local symmetric iid case. For each code, we apply an optimal SDP recovery map~\cite{fletcher2007optimum}, providing strict lower bound on the entanglement infidelities. Therefore, the results provide a metric to experimentalists for selecting short-length PI codes for near-term implementation. We also emphasize that the Barnum--Knill/Petz recovery is near-optimal, with the entanglement infidelity bounded within a factor of two of the optimum~\cite{barnum2002reversing}.



For global symmetric correlated AD noise, the CAD PI codes show a clear advantage over other short-length PI codes considered here. While CAD4 already performs competitively, CAD9 gives the strongest performance across the full range of AD probabilities and, at small noise strengths, suppresses the entanglement infidelity by several orders of magnitude compared with the second best performing 7-qubit AAB code. This highlights the benefit of designing bespoke codes.  

For local symmetric, or iid, AD noise, the trend is different. In this case, the \(((7,1,3))\)-AAB code gives the best overall performance among the codes tested, with the other distance-\(3\) PI codes following closely. The CAD codes, especially CAD9, no longer perform as well in this iid setting. This reflects that these codes are optimized for correlated AD errors and are therefore not expected to be good against iid AD noise. When the noise becomes local and less structured, good performance is closer to the usual distance-based condition for arbitrary errors. 

Another key observation is that, for iid AD noise, larger codes with the same distance tend to perform worse than smaller codes. However, the trend does not carry over to the global symmetric case. For small \(p\), the \(((9,1,3))\)-bgm code and the \(((11,1,3))\) Kubischta--Teixeira outperform the 7-qubit Pollatsek--Ruskai code, showing that the choice of PI codes depends strongly on whether the AD noise is iid or correlated.

\subsection{Spatially correlated errors}\label{sec:spatially correlated errors}

We now numerically show the effectiveness of QER against correlated AD errors. We again use the \(((9,1,3))\)-bgm code, evolve under the exact global symmetric correlated AD channel. While the Petz recovery map is constructed only from truncated first- and second-order AD error terms. For comparison, we also plot the SDP-optimized recovery for full channel and the bare-qubit case.

Fig.~\ref{fig:penalty} shows that the second-order Petz recovery (orange) closely follows the SDP-optimized recovery (purple). This indicates that the dominant errors of the exact channel is already captured by keeping terms up to second order. In contrast, the first-order Petz recovery performs several orders of magnitude worse, showing that weight-two, or spatially correlated AD errors, cannot be ignored. In fact, they become the leading uncorrected contribution once the first-order errors are recovered. 

This result also supports the practical feasibility of implementing a truncated recovery map. Since the second-order recovery already approaches the SDP optimum, one does not need to include the full Kraus expansion of the exact channel. This reduces the number of Kraus operators, and therefore the number of ancillas and time steps required by the recovery algorithm, with only a small penalty in performance.

\section{Discussion}\label{sec:discussion}

In this work, we bridged the gap between abstract quantum error recovery maps and their hardware implementation by compiling the coherent CPTP-map implementation protocol of Ticozzi and Viola~\cite{ticozzi2017quantum} for recovery maps acting on PI codes. The compiled gates are written in terms of Dicke-space primitives, which can be realized using linear or nonlinear GPGs. We also tested the compiled recovery circuit with a noisy GPG model and showed that the recovery advantage survives realistic control imperfections. By exploiting the structure of global symmetric correlated AD noise, we introduced a new family of PI codes, called CAD codes, which perform several orders of magnitude better than the other short-length PI codes studied here. Our numerical results also show that correlated and iid AD noise affect PI codes very differently, highlighting the importance of matching the code and recovery strategy to the dominant physical error model. 

Although we focus on AD noise here, our accompanying GitHub repository \cite{chandra_qer_pi_2026} can handle depolarizing noise as well as customized error models. This makes it straightforward to extend the study of PI codes to other noise models by modifying the error channel in the repository.   

QER typically requires some knowledge of the error channel. This is a reasonable assumption in many hardware settings, where the dominant noise mechanism can often be identified experimentally or by device modeling. For example, energy relaxation or \(T_1\)-limited AD noise is a central limitation in superconducting circuits~\cite{kubica2023erasure,klimov2018fluctuations}, atom loss is a major issue in neutral atoms \cite{chow2024circuit, wu2022erasure} and ion traps \cite{kang2023quantum}. In such settings, one can use knowledge of the dominant physical error process and use QER schemes for better error suppression than usual QEC.

This work opens several directions: benchmarking larger-distance CAD PI codes, \(\mathrm{bgm}\) codes and so on. Another natural direction is to use small PI codes like CAD4 code as an inner layer concatenation inside a stabilizer code, so that the AD errors get suppressed before the noise reaches the outer code. Finally, demonstrating this protocol in platforms with native collective control, such as neutral atoms in optical/microwave cavities or trapped ions, would be an important step toward practical low-overhead recovery of AD errors and demonstration of QEC.

\section{Acknowledgment}
O.C. would like to thank Daniel Madden for insightful discussions during the early phase of this project. 
O.C. is supported by Sydney Quantum Academy, Sydney, Australia. 
Y.O. acknowledges support from EPSRC Grant No. EP/W028115/1 and also the EPSRC funded QCI3 Hub under Grant No. EP/Z53318X/1.
\bibliographystyle{unsrt}
\bibliography{library}

\onecolumngrid
\appendix
\section{Permutation-Invariant States and the Heisenberg Ferromagnet}
\label{app: pi states and heisenberg}

Permutation-invariant quantum codes (PI codes) fundamentally reside in the ground space of a Heisenberg ferromagnetic Hamiltonian~\cite{ouyang2021quantum}. A ferromagnetic system can be modeled as an ensemble of spin-\(1/2\) particles interacting with one another in the absence of an external magnetic field. The Hamiltonian for this system is given by
\begin{equation} \label{eq: heisenberg_hamiltonian}
    H = -2\sum_{i<j}J_{ij}S_{i}\cdot S_{j},
\end{equation}
where \(J_{ij} > 0\) represents the exchange interaction coupling (positive for a ferromagnetic model), and \(S_{i}\) and \(S_{j}\) are the spin operators of the \(i\)-th and \(j\)-th particles, respectively. To see how this spin interaction relates to the permutation operator \(P_{ij}\), which swaps the states of spin-\(i\) and spin-\(j\), consider the combined spin operator \(S_{ij} = S_i + S_j\). Squaring both sides yields:
\begin{align}
    S_{ij}^2 &= (S_i + S_j)^2 \nonumber \\
             &= S_i^2 + S_j^2 + S_i \cdot S_j + S_j \cdot S_i.
\end{align}
Since \([S_i, S_j] = 0\) for all \(i \neq j\), we have \(S_{ij}^2 = S_i^2 + S_j^2 + 2S_i \cdot S_j\), which rearranges to:
\begin{equation}
    2S_i \cdot S_j = S_{ij}^2 - S_i^2 - S_j^2.
\end{equation}
Recall that \(S_i^2 \vert J, m \rangle = J(J+1) \vert J, m \rangle\) in the angular momentum basis. For spin-\(1/2\) particles, \(J = 1/2\), meaning the eigenvalues of both \(S_i^2\) and \(S_j^2\) are \(3/4\). Furthermore, the combination of two spin-\(1/2\) particles decomposes into irreducible representations of \(SU(2)\) as \(1/2 \otimes 1/2 = 0 \oplus 1\). This implies that the total spin \(s\) can be \(0\) or \(1\), corresponding to eigenvalues of \(S_{ij}^2\) equal to \(0\) and \(2\), respectively. Hence, the eigenvalues of \(S_i \cdot S_j\) are:
\begin{equation}
    S_i \cdot S_j = \begin{cases} 
        1/4, & \text{for } S_{ij}^2 = 2 \ (s=1), \\ 
        -3/4, & \text{for } S_{ij}^2 = 0 \ (s=0). 
    \end{cases}
\end{equation}
Note that \(s=1\) corresponds to the symmetric subspace, and \(s=0\) corresponds to the antisymmetric subspace. Let \(\Pi_s\) and \(\Pi_{as}\) denote the projectors onto the symmetric and antisymmetric subspaces, respectively. We can rewrite the operator \(S_i \cdot S_j\) as:
\begin{equation}
    S_i \cdot S_j = \frac{1}{4}\Pi_s - \frac{3}{4}\Pi_{as}.
\end{equation}
Since \(\Pi_s + \Pi_{as} = I\), we can express the projectors as:
\begin{align}
    \Pi_s &= S_i \cdot S_j + \frac{3}{4}I, \\ \nonumber
    \Pi_{as} &= \frac{1}{4}I - S_i \cdot S_j.
\end{align}
The permutation operator \(P_{ij}\) leaves the symmetric space unchanged (eigenvalue \(+1\)) and negates the antisymmetric space (eigenvalue \(-1\)). Thus, we can write \(P_{ij}\) in terms of these projectors:
\begin{align}
    P_{ij} &= \Pi_s - \Pi_{as} \nonumber \\
           &= \left(S_i \cdot S_j + \frac{3}{4}I\right) - \left(\frac{1}{4}I - S_i \cdot S_j\right) \nonumber \\
           &= \frac{I}{2} + 2S_i \cdot S_j. \label{eq: permutation operator}
\end{align}

Using Eq.~\ref{eq: permutation operator}, the Hamiltonian can be rewritten entirely in terms of permutation operators:
\begin{equation}
    H = -\sum_{i<j}J_{ij}\left(P_{ij} - \frac{I}{2}\right).
\end{equation}

A state \(\vert \Psi \rangle\) is defined as permutation-invariant if it lies in the \(+1\) eigenspace of all permutation operators \(P_{ij}\), such that \(P_{ij}\vert \Psi \rangle = \vert \Psi \rangle\) for all \(i < j\). For simplicity, let us denote the total exchange coupling as \(J = \sum_{i<j}J_{ij}\). If we apply the shifted Hamiltonian \((H - \frac{J}{2}I)\) to a permutation-invariant state \(\vert \Psi \rangle\), we find:
\begin{align}
    \left(H - \frac{J}{2}I\right)\vert \Psi \rangle &= -\sum_{i<j}J_{ij}P_{ij}\vert \Psi \rangle \nonumber \\
    &= -J\vert \Psi \rangle,
\end{align}
which implies that \(H\vert \Psi \rangle = -\frac{J}{2}\vert \Psi \rangle\). 

To demonstrate that this eigenvalue represents the ground state of the system, we can bound the operator norm of the shifted Hamiltonian. By applying the triangle inequality, we obtain:
\begin{align}
    \left\|H - \frac{J}{2}I\right\| &= \left\|-\sum_{i<j}J_{ij}P_{ij}\right\| \nonumber \\
    &\le \sum_{i<j}J_{ij}\|P_{ij}\| \nonumber \\
    &= \sum_{i<j}J_{ij} \nonumber \\
    &= J.
\end{align}
Because \((H - \frac{J}{2}I)\) is a Hermitian operator and its norm is bounded by \(J\), its eigenvalues must lie in the interval \([-J, J]\). Consequently, the eigenvalues of the original Hamiltonian \(H\) are restricted to the interval \(\left[-\frac{J}{2}, \frac{3J}{2}\right]\). 

Since permutation-invariant states \(\vert \Psi \rangle\) achieve the absolute minimum possible eigenvalue of \(-\frac{J}{2}\), we conclude that all PI states, and therefore PI codes, belong to the exact ground space of a Heisenberg ferromagnetic system.

\section{Collective basis representation} \label{app: collective basis representation}
If the assumption that spins are closely spaced as compared to the wavelength of the environment mode does not holds then Hamiltonian in eq.~\eqref{eq:interaction hamiltonian} is position dependent. Which means that the modes acts on every spin uniquely. Implying that now we cannot use the collective operators which enormously helped us reduce the dimensionality. However, we can use collective representation of density matrix characterized by different total angular momentum \(J\) as done in \cite{chase2008collective, baragiola2010collective} to study the effect. 

Consider a system of \(N\) identical spin-\(1/2\)s. The joint Hilbert space for the entire spin ensemble is \(\mathcal{H} = \mathcal{H}^1 \otimes \mathcal{H}^2 ...\otimes \mathcal{H}^N\). A pure state of the ensemble can be written as 
\begin{align}
    \ket{\Psi} = \sum_{\vec{m} =\{m_1,m_2,..,m_N\}} c_{\vec{m}} \ket{m_1, m_2,...,m_N},
\end{align}
where the basis \(\ket{m_1, m_2,...,m_N}\) is a simultaneous eigenstate of \(J^2\) and \(J_z\) with eigenvalues given by 
\begin{align}
J^2 \ket{m_1, m_2,\dots,m_N}
&= J(J+1)\,\ket{m_1, m_2,\dots,m_N}, \\
J_z \ket{m_1, m_2,\dots,m_N}
&= \left(\sum_j m_j\right)\,\ket{m_1, m_2,\dots,m_N},
\end{align}
 
Each particle undergoes through an operation \(U\) where \(U \in SU(2)\) group. The combined operation \(\mathbf{U} = U_1 \otimes U_2 \otimes ...\otimes U_N\) belongs to \(SU(2)^N\) group. The special case when operations acting on every spin is identical i.e., \(U_1=U_2=..=U_N\), the combined operation \(\mathbf{U} = U^N\) becomes reducible and can be decomposed into a direct sum of irreducible spin-\(J\) sectors, labeled by the total angular momentum \(J\) and its magnetic quantum number \(M\), with additional multiplicities reflecting the different ways a given \(J\) arises from adding \(N\) spin-\(1/2\) particles. Accordingly, the Hilbert space admits the Schur-Weyl decomposition \((C^2)^N \cong \oplus _J (V_J \otimes M_J)\), where \(V_J\) is the \(2J+1\)-dimensional irrep carrying the collective spin-J action and \(M_J\) is a multiplicity space of dimension \(d_N^J\) on which collective rotations act trivially. We can write the combined operation \(\bold{U}\) as 
\begin{align}
\mathbf{U}
= \bigoplus_{J=J_{\min}}^{J_{\max}}
\left(
  \bigoplus_{i=1}^{d_{N}^{J}} U^{J,i}
\right),
\end{align}
where \(d_N^J\) is degeneracy within each \(J\) given by \cite{mikhailov1977addition}
\begin{align}
d_{N}^{J}
&= \frac{N!\,(2J+1)}{\left(\frac{N}{2}-J\right)!\left(\frac{N}{2}+J+1\right)!}\,,   
\end{align}
\begin{align*}
J_{\min} &=
\begin{cases}
\frac{1}{2}, & \text{if \(N\) is odd},\\[4pt]
0, & \text{if \(N\) is even},
\end{cases}
\qquad
J_{\max} = \frac{N}{2}.
\end{align*}

Since the system evolves under an operation \(\mathbf{U}\) whose action decomposes into distinct \(J\) sectors, the density matrix can be expressed in the same block-diagonal structure. We refer to this as the collective density matrix \(\rho_c\), defined as the direct sum over the \(J\)-irrep blocks,
\begin{align}
    \rho_c = \bigoplus_J \rho_J = \sum_J \sum_{M,M'} c_{J,M,M'} \ket{J,M}\bra{J,M'}, 
\end{align}
where \(c_{J,M,M'}\) denotes coefficients. 
Under the                                    operations of the form \(\mathbf{U}\) (i.e., local-symmetric maps), a super-selection rule applies: these maps cannot create coherence between different irrep blocks \cite{stockton2003characterizing}. Consequently, the dynamics only mix different \(M\) values within a fixed \(J\) sector, and allow transitions only between neighboring spin sectors, i.e., \(J \to J\pm 1\). 

\section{Amplitude damping channel in vectorized picture} \label{appendix: global symmetric noise}
\subsection{Vectorization}
A system \(\rho\) starts in the codespace of a PI code which can be written as
\begin{equation}
    \rho = \sum_{m,n} \rho_{m,n} \ket{J = \frac{N}{2}, J_z = m  } \bra{J = \frac{N}{2}, J_z = n}.
\end{equation}
In the vectorized form \(\rho\) becomes
\begin{equation}
    |\rho \rangle \rangle = \sum_{m,n} \rho_{m,n} \ket{m} \otimes \ket{n}, 
\end{equation}
where we have omitted writing the value of \(J\) because it is \(N/2\) for Dicke subspace. The Lindbladian in the vectorized form becomes, 
\begin{equation} 
    \mathrm{L} = \sum_{i} (A_i \otimes A_i^* - \frac{1}{2} (A_i^{\dagger}A_i \otimes I + I \otimes A_i^{\dagger} A_i) ), 
\end{equation}
where \(A_i\) are the jump operators. The time evolution equation is, \begin{align}
    \frac{d}{dt} |\rho \rangle \rangle & =  \mathrm{L} |\rho \rangle \rangle \\ 
    |\rho(t) \rangle \rangle & =  \mathrm{S} |\rho (0) \rangle \rangle,  
\end{align}
where \(\mathrm{S}(t)  = e^{\mathrm{L}t}\) is the superoperator. 

\subsection{Global symmetric amplitude damping} \label{appendix: closed form derivation of AD channel in doubled-space}

The vectorized Lindbladian for the global symmetric  amplitude damping channel is 
\begin{equation} \label{eq: lindbladian for amplitude damping in doubled space}
    \mathrm{L} =  \gamma_- \left( J_- \otimes J_- - \frac{1}{2} (J_-^{\dagger}J_- \otimes I + I \otimes J_-^{\dagger} J_-) \right ). 
\end{equation}
Now the action of \(\mathrm{L}\) on the doubled-space \(|m\rangle \otimes |n\rangle\) introduced above is, 
\begin{align} \label{eq:action of the linear L in doubled space}
    \mathrm{L}|m,n \rangle \rangle = \sqrt{\gamma_m \gamma_n} |m-1, n-1 \rangle \rangle - \beta_0 ^{(m,n)} |m, n \rangle \rangle,  
\end{align}
and the coefficients are,
\begin{align}
\beta_{r}^{(m,n)} &= \frac{\gamma_{m-r} + \gamma_{n-r}}{2} \\
\gamma_{m-r} &= (J+m-r)\bigl(J-m+r+1\bigr),
\end{align}
where \(r\) denotes the number of excitation loss \(|m-r, n-r \rangle \rangle\). For the above case, \(r=0\). 

Let $k_{\max} = \min\!\left(m,\, n\right)$. Now \(J_-\) can act until \(k_{max}\) times before hitting the lowermost state after which the eigenvalue is \(0\). That means the exponential is a finite sum 
\begin{align}
    e^{\mathrm{L}t} |m, n\rangle \rangle = \sum_{k=0} ^{k_{max}} C_k^{(m,n)} (t) |m-k, n-k \rangle \rangle, 
\end{align}
The time evolution equation in this compact form is,
\begin{align}
    | \rho (t) \rangle \rangle = e^{\mathrm{L}t} | \rho (0) \rangle \rangle = \sum_{k=0}  ^{k_{max}} C_k^{(m,n)}(t) | m-k, n-k \rangle \rangle.  
\end{align}
with the initial condition \(C_0 (0) = 1\) and \(C_{k \geq 1} (0) = 0\). Meaning that at time \(t=0\) the state starts in \(|m ,n \rangle \rangle
\) and the occupation of all other states is \(0\). We know, 
\begin{align}
\frac{d}{dt}|\rho(t) \rangle \rangle = \mathrm{L} |\rho(t) \rangle \rangle.
\end{align}
From the LHS of the equation we get, 
\begin{align}
\frac{d}{dt} |\rho (t) \rangle \rangle  = \sum_{k=0} ^ {k_{max}} \dot{C}_k ^{(m,n)} (t) |m -k, n-k  \rangle \rangle,   
\end{align}
and from the RHS we get
\begin{align}
\mathrm{L} \lvert \rho(t) \rangle\rangle
&= \sum_{k=0}^{k_{max}} C_k ^{(m,n)}(t)
\Big(
\alpha_{k+1}
\lvert m-(k+1),\, n-(k+1) \rangle\rangle
\nonumber \\
&\quad
- \beta_k
\lvert m-k,\, n-k \rangle\rangle
\Big),
\end{align}
where
\begin{equation}
\beta_k = \frac{\gamma_{m-k} + \gamma_{n-k}}{2}
\quad \text{and} \quad
\alpha_{k+1} = \sqrt{\gamma_{m-k}\gamma_{n-k}} .
\end{equation} 
Now compare the LHS and RHS for \(k=0\) and get, 
\begin{align} \label{eq: C_0(t) in time domain}
\dot{C}_0^{(m,n)}(t) &= -\beta_0 C_0 ^{(m,n)}(t), \\
\intertext{which can be immediately solved using the initial condition \(C_0 ^{(m,n)}(0)=1\) to give}
C_0 ^{(m,n)}(t) &= e^{-\beta_0 t}.
\end{align}
Now for \(k \geq 1\) we get, 
\begin{align} \label{eq: equation for k >=1}
    \dot{C}_k ^{(m,n)} (t) = -\beta_k C_k^{(m,n)} + \alpha_{k} C_{k-1}^{(m,n)}.   
\end{align}
In order to convert the above partial differential equation into a polynomial equation we will use Laplace transform. Let's first introduce the Laplace transform 
\begin{align}
    \tilde{C}_k ^{(m,n)} (p) = \mathcal{L} \{C_k^{(m,n)}(t)\} =  \int_0 ^ {\infty} e^{-pt} C_k^{(m,n)}(t) dt.
\end{align}
An important identity we will be using is 
\begin{align}
    \mathcal L\{\dot C_k ^{(m,n)}\}=p\widetilde C_k^{(m,n)}(p)-C_k^{(m,n)}(0).
\end{align}

Now applying Laplace transform to equation \ref{eq: equation for k >=1} gives us, 
\begin{align}
    (p + \beta_k) \tilde{C}_k ^{(m,n)}(p)& = \alpha_k \tilde{C}_{k-1}^{(m,n)}(p) \\ 
    \tilde{C}_k^{(m,n)}(p) & = \frac{\alpha_k \tilde{C}_{k-1}^{(m,n)}(p)}{p+\beta_k}, 
\end{align}
using the condition \(C_{k \geq 1}^{(m,n)} (0) =0\). 
We get an iterating equation for every \(k\) where the value of \(\tilde{C}_k^{(m,n)}(p)\) depends on \(\tilde{C}_{k-1}^{(m,n)}(p)\). We can keep iterating to get a closed form expression in Laplace space 
\begin{equation} \label{eq: C_k in laplace's space}
    \widetilde C_k ^{(m,n)}(p)
    \;=\;
    \frac{\displaystyle\prod_{r=1}^{k}\alpha_r}{\displaystyle\prod_{r=0}^{k}(p+\beta_r)},
\end{equation}

where \(\tilde{C}_0^{(m,n)}(p) = \frac{1}{p+\beta_0}\) is derived from applying Laplace transform to equation \ref{eq: C_0(t) in time domain}. \newline

Now in order to bring the equation back to variable \(t\) (time), we will apply inverse Laplace transform on~\eqref{eq: C_k in laplace's space}. 
Assume first that $\beta_0,\beta_1,\dots,\beta_k$ are pairwise distinct. We expand Eq.~\eqref{eq: C_k in laplace's space} into partial fractions,
\begin{equation}
    \frac{1}{\prod_{r=0}^{k}(p+\beta_r)}
    \;=\;
    \sum_{q=0}^{k}\frac{A_q}{p+\beta_q},
    \label{eq:partial_fractions_app}
\end{equation}
where the coefficients are obtained by evaluation at the poles:
\begin{equation}
    A_q
    \;=\;
    \left.\frac{1}{\prod_{r\neq q}(p+\beta_r)}\right|_{p=-\beta_q}
    \;=\;
    \frac{1}{\prod_{r\neq q}(\beta_r-\beta_q)}.
    \label{eq:Aq_app}
\end{equation}
Using $\mathcal L^{-1}\{(p+\beta_q)^{-1}\}=e^{-\beta_q t}$, we obtain
\begin{equation} \label{eq: final coeficients of action of channel in doubled space}
    C_k^{(m,n)}(t)
    \;=\;
    \Bigl(\prod_{r=1}^{k}\alpha_r^{(m,n)}\Bigr)
    \sum_{q=0}^{k}
    \frac{e^{-\beta_q^{(m,n)} t}}{\prod_{r\neq q}\Bigr(\beta_r^{(m,n)}-\beta_q^{(m,n)}\Bigl)},
\end{equation}
with \(\alpha_r^{(m,n)} = \sqrt{\gamma_{m-r+1} \gamma_{n-r+1}}\) and \(\beta_q^{(m,n)} = \frac{\gamma_{m-q} + \gamma_{n-q} }{2}\). \newline

Note that if some $\beta_r$ coincide, then Eq.~\eqref{eq: C_k in laplace's space} has repeated poles and the inverse Laplace transform
produces terms of the form $t^{\ell}e^{-\beta t}$ (with $\ell$ determined by the multiplicity). 
These expressions can be obtained by taking the appropriate limit as the
colliding $\beta$'s approach one another, or by performing partial fractions with repeated poles directly.

Finally, we get the action of the channel \(S = e^{\mathrm{L}t}\) in the doubled space, 
\begin{equation}
S_t|m,n\rangle \rangle  =   e^{\mathrm{L}t}\,|m, n \rangle \rangle 
    \;=\;
    \sum_{k=0}^{k_{max}} C_k^{(m,n)}(t)\,|m-k,n-k \rangle \rangle,
\end{equation} with coefficients \(C_k ^{(m,n)}\) given in Eq. \ref{eq: final coeficients of action of channel in doubled space}. 

\subsection{Truncated global symmetric AD channel}\label{app:truncated global symmetric AD} 

\subsubsection{Up to first order in \(t\)} \label{app:truncated global symmetric AD upto first order}

We now derive a Kraus representation for the first-order in $t$ truncation of the global symmetric amplitude-damping channel. The first-order approximation to the vectorized evolution is
\begin{align}
S_1 = e^{\mathrm{L}t} \simeq I \otimes I + \mathrm{L}t .
\end{align}
Using the action of $\mathrm{L}$ on the doubled basis derived above, it is clear that $S_1$ only connects
\begin{align}
|m,n\rangle\rangle \longrightarrow |m,n\rangle\rangle, \qquad |m-1,n-1\rangle\rangle .
\end{align}
We therefore make the first-order Kraus ansatz
\begin{align}
\widetilde{K}_0 |m\rangle = d_m |m\rangle , \qquad \widetilde{K}_1 |m\rangle = s_m |m-1\rangle .
\label{eq:fo_kraus_ansatz}
\end{align}
Then, acting on the operator basis element $|m\rangle\langle n| \equiv |m,n\rangle\rangle$, we get
\begin{align}
\widetilde{K}_0 |m\rangle\langle n| \widetilde{K}_0^\dagger &= d_m d_n |m,n\rangle\rangle , \\
\widetilde{K}_1 |m\rangle\langle n| \widetilde{K}_1^\dagger &= s_m s_n |m-1,n-1\rangle\rangle .
\end{align}
Hence,
\begin{align}
\mathcal{E}^{(1)}_t \left( |m,n\rangle\rangle \right) = d_m d_n |m,n\rangle\rangle + s_m s_n |m-1,n-1\rangle\rangle + \mathcal{O}((\gamma_- t)^2).
\label{eq:fo_kraus_action}
\end{align}
Comparing Eq.~\eqref{eq:fo_kraus_action} with Eq.~\eqref{eq:action of the linear L in doubled space}, we identify
\begin{align}
d_m d_n &= 1 - \frac{\gamma_- t}{2}\left(\gamma_m + \gamma_n\right) + \mathcal{O}((\gamma_- t)^2), \label{eq:fo_dm_dn_match} \\
s_m s_n &= \gamma_- t \sqrt{\gamma_m\gamma_n}. \label{eq:fo_sm_sn_match}
\end{align}
A convenient choice satisfying these relations is
\begin{align}
d_m &= 1 - \frac{\gamma_- t}{2}\gamma_m , \label{eq:fo_dm_def} \\
s_m &= \sqrt{\gamma_- t \gamma_m}. \label{eq:fo_sm_def}
\end{align}
Indeed,
\begin{align}
d_m d_n &= \left( 1 - \frac{\gamma_- t}{2}\gamma_m \right) \left( 1 - \frac{\gamma_- t}{2}\gamma_n \right) \nonumber \\
&= 1 - \frac{\gamma_- t}{2}\left(\gamma_m + \gamma_n\right) + \mathcal{O}((\gamma_- t)^2), \\
s_m s_n &= \sqrt{\gamma_- t \gamma_m}\sqrt{\gamma_- t \gamma_n} = \gamma_- t \sqrt{\gamma_m\gamma_n}.
\end{align}
Therefore, the first-order Kraus operators act on the Dicke basis as
\begin{align}
\widetilde{K}_0 |m\rangle &= \left( 1 - \frac{\gamma_- t}{2}\gamma_m \right) |m\rangle , \\
\widetilde{K}_1 |m\rangle &= \sqrt{\gamma_- t \gamma_m} |m-1\rangle .
\end{align}
Equivalently, in operator form,
\begin{align}
\widetilde{K}_0 &= I - \frac{\gamma_- t}{2} J_+ J_- , \label{eq:fo_two_kraus_k0} \\
\widetilde{K}_1 &= \sqrt{\gamma_- t} J_- . \label{eq:fo_two_kraus_k1}
\end{align}
The final completed first-order Kraus including the third term to satisfy the completeness condition is therefore
\begin{align}
K^{(1)}_0 &= \left[ I - \gamma_- t J_+ J_- - \frac{(\gamma_- t)^2}{4} \left(J_+ J_-\right)^2 \right]^{1/2}, \label{eq:fo_final_k0} \\
K^{(1)}_1 &= \sqrt{\gamma_- t} J_- , \label{eq:fo_final_k1} \\
K^{(1)}_2 &= \frac{\gamma_- t}{2} J_+ J_- . \label{eq:fo_final_k2}
\end{align}

\subsubsection{Up to second order in \(t\) } \label{app:truncated global symmetric AD upto second order}
Now we will see the effect of the channel approximated to the second order in $t$: 
\begin{align}
   S_2 = e^{\mathrm{L}t} \approx \mathbb{I} \otimes \mathbb{I} + \frac{\mathrm{L}t}{1!} + \frac{(\mathrm{L}t)^2}{2!}, 
\end{align}
where \(\mathrm{L}\) is the Lindbladian for AD channel in doubled space as defined in Eq.~\ref{eq: lindbladian for amplitude damping in doubled space}. 
Define
\begin{equation}
N := J_+J_-,
\qquad
A := J_- \otimes J_-,
\qquad
B := \frac{1}{2}\bigl(N\otimes \mathbb{I} + \mathbb{I}\otimes N\bigr),
\qquad
L = \gamma_-(A-B),
\end{equation}
and
\begin{align}
\mathrm{L}^{2}
=
\gamma_{-}^{2}(A-B)^{2}
=
\gamma_{-}^{2}\bigl(A^{2}-AB-BA+B^{2}\bigr),
\end{align}
we compute each term separately:
\begin{equation}
\begin{aligned}
A^{2} &:= J_-^{2}\otimes J_-^{2}, \qquad
AB := \frac{1}{2}\bigl(J_-N\otimes J_- + J_-\otimes J_-N\bigr),\\
BA &:= \frac{1}{2}\bigl(NJ_-\otimes J_- + J_-\otimes NJ_-\bigr), \qquad
B^{2} := \frac{1}{4}\bigl(N^{2}\otimes \mathbb{I}
+2N\otimes N+\mathbb{I}\otimes N^{2}\bigr).
\end{aligned}
\end{equation}
Recall that we already know the action of \(\mathrm{L}\) \eqref{eq:action of the linear L in doubled space}. The action of \(\mathrm{L}^2\) term is, 
\begin{equation}
\begin{aligned}
\mathrm{L}^2 |m,n\rangle\!\rangle
= \gamma_-^2 \biggl[
&\sqrt{\gamma_m\gamma_n\gamma_{m-1}\gamma_{n-1}}\,|m-2,n-2\rangle\!\rangle
-\frac{1}{2}\sqrt{\gamma_m\gamma_n}\,
\bigl(\gamma_m+\gamma_n+\gamma_{m-1}+\gamma_{n-1}\bigr)
|m-1,n-1\rangle\!\rangle \\
&+\frac{1}{4}\bigl(\gamma_m+\gamma_n\bigr)^2
|m,n\rangle\!\rangle
\biggr].
\end{aligned}
\end{equation}

Therefore, the final expression becomes in compact form is,
\begin{align}
e^{\mathrm{L}t}|m,n\rangle\!\rangle
=
A_{m,n}(t)\,|m,n\rangle\!\rangle
+
B_{m,n}(t)\,|m-1,n-1\rangle\!\rangle
+
C_{m,n}(t)\,|m-2,n-2\rangle\!\rangle
+
\mathcal{O}(t^3),
\label{eq:truncated_map_on_doubled_basis}
\end{align}
where
\begin{align}
A_{m,n}(t)
&= \left( 1
-\frac{\gamma_- t}{2}(\gamma_m+\gamma_n)
+\frac{\gamma_-^2 t^2}{8}(\gamma_m+\gamma_n)^2
\right),
\\
B_{m,n}(t)
&=\left(
\gamma_- t\,\sqrt{\gamma_m \gamma_n}
-\frac{\gamma_-^2 t^2}{2} \sqrt{\gamma_m \gamma_n}
\left(
\frac{\gamma_m+\gamma_n}{2}
+
\frac{\gamma_{m-1}+\gamma_{n-1}}{2}
\right)
\right),
\\
C_{m,n}(t)
&=
\frac{\gamma_-^2 t^2}{2}\,
\sqrt{\gamma_m \gamma_n \gamma_{m-1} \gamma_{n-1}}.
\end{align}

Clearly, the truncated map only connects
\begin{align}
|m,n\rangle\!\rangle
\;\longrightarrow\;
|m,n\rangle\!\rangle,\quad
|m-1,n-1\rangle\!\rangle,\quad
|m-2,n-2\rangle\!\rangle
\end{align}
up to second order in \(t\). We therefore make the ansatz
\begin{equation}
\label{eq:K_ansatz}
K_0|m\rangle=d_m|m\rangle,\qquad
K_1|m\rangle=s_m|m-1\rangle,\qquad
K_2|m\rangle=u_m|m-2\rangle.
\end{equation}
Then, acting on the operator basis element \(|m\rangle\langle n| \equiv |m,n\rangle\!\rangle\),
\begin{align}
K_0 |m\rangle\langle n| K_0^\dagger
&=
d_m d_n\, |m,n\rangle\!\rangle,
\\
K_1 |m\rangle\langle n| K_1^\dagger
&=
s_m s_n\, |m-1,n-1\rangle\!\rangle,
\\
K_2 |m\rangle\langle n| K_2^\dagger
&=
u_m u_n\, |m-2,n-2\rangle\!\rangle.
\end{align}
Hence
\begin{align}
\mathcal{E}^{(2)}_t\bigl(|m,n\rangle\!\rangle\bigr)
=
d_m d_n\, |m,n\rangle\!\rangle
+
s_m s_n\, |m-1,n-1\rangle\!\rangle
+
u_m u_n\, |m-2,n-2\rangle\!\rangle
+
\mathcal{O}(t^3).
\label{eq:kraus_action_operator_basis}
\end{align}

Comparing Eq.~\eqref{eq:kraus_action_operator_basis} with Eq.~\eqref{eq:truncated_map_on_doubled_basis}, we identify
\begin{align}
d_m d_n
&=
1
-\frac{\gamma_- t}{2}(\gamma_m+\gamma_n)
+\frac{\gamma_-^2 t^2}{8}(\gamma_m+\gamma_n)^2,
\label{eq:dm_dn_match}
\\
s_m s_n
&=
\gamma_- t\, \sqrt{\gamma_m\gamma_n}
-\frac{\gamma_-^2 t^2}{2}\sqrt{\gamma_m\gamma_n}
\left(
\frac{\gamma_m+\gamma_n}{2}
+
\frac{\gamma_{m-1}+\gamma_{n-1}}{2}
\right),
\label{eq:sm_sn_match}
\\
u_m u_n
&=
\frac{\gamma_-^2 t^2}{2}\,
\sqrt{\gamma_m\gamma_n\gamma_{m-1}\gamma_{n-1}}.
\label{eq:um_un_match}
\end{align}

A convenient choice satisfying these relations is
\begin{align}
d_m
&=
1-\frac{\gamma_- t}{2}\gamma_m+\frac{\gamma_-^2 t^2}{8}\gamma_m^2,
\label{eq:dm_def}
\\
s_m
&=
\sqrt{\gamma_- t\,\gamma_m}
\left[
1-\frac{\gamma_- t}{4}\bigl(\gamma_m+\gamma_{m-1}\bigr)
\right],
\label{eq:sm_def}
\\
u_m
&=
\frac{\gamma_- t}{\sqrt{2}}\sqrt{\gamma_m\gamma_{m-1}}.
\label{eq:um_def}
\end{align}
Indeed,
\begin{align}
d_m d_n
&=
\left(
1-\frac{\gamma_- t}{2}\gamma_m+\frac{\gamma_-^2 t^2}{8}\gamma_m^2
\right)
\left(
1-\frac{\gamma_- t}{2}\gamma_n+\frac{\gamma_-^2 t^2}{8}\gamma_n^2
\right)
\nonumber\\
&=
1-\frac{\gamma_- t}{2}(\gamma_m+\gamma_n)
+\frac{\gamma_-^2 t^2}{8}(\gamma_m+\gamma_n)^2
+\mathcal{O}(t^3),
\\[1mm]
s_m s_n
&=
\sqrt{\gamma_- t\,\gamma_m}
\left[
1-\frac{\gamma_- t}{4}(\gamma_m+\gamma_{m-1})
\right]
\sqrt{\gamma_- t\,\gamma_n}
\left[
1-\frac{\gamma_- t}{4}(\gamma_n+\gamma_{n-1})
\right]
\nonumber\\
&=
\gamma_- t\,\sqrt{\gamma_m\gamma_n}
-\frac{\gamma_-^2 t^2}{4}\sqrt{\gamma_m\gamma_n}
\bigl(
\gamma_m+\gamma_{m-1}+\gamma_n+\gamma_{n-1}
\bigr)
+\mathcal{O}(t^3)
\nonumber\\
&=
\gamma_- t\,\sqrt{\gamma_m\gamma_n}
-\frac{\gamma_-^2 t^2}{2}\sqrt{\gamma_m\gamma_n}
\left(
\frac{\gamma_m+\gamma_n}{2}
+
\frac{\gamma_{m-1}+\gamma_{n-1}}{2}
\right)
+\mathcal{O}(t^3),
\\[1mm]
u_m u_n
&=
\frac{\gamma_-^2 t^2}{2}\,
\sqrt{\gamma_m\gamma_n\gamma_{m-1}\gamma_{n-1}}.
\end{align}

Therefore, the Kraus operators act on the Dicke basis as
\begin{align}
K_0 |m\rangle
&=
\left(
1-\frac{\gamma_- t}{2}\gamma_m+\frac{\gamma_-^2 t^2}{8}\gamma_m^2
\right)|m\rangle,
\\
K_1 |m\rangle
&=
\sqrt{\gamma_- t\,\gamma_m}
\left[
1-\frac{\gamma_- t}{4}\bigl(\gamma_m+\gamma_{m-1}\bigr)
\right]
|m-1\rangle,
\\
K_2 |m\rangle
&=
\frac{\gamma_- t}{\sqrt{2}}\sqrt{\gamma_m\gamma_{m-1}}\, |m-2\rangle.
\end{align}

Equivalently, in operator form,
\begin{align} 
K_0
&=
\mathbb{I}
-\frac{\gamma_- t}{2} J_+J_-
+\frac{\gamma_-^2 t^2}{8} (J_+J_-)^2,
\label{eq:K0_final}
\\
K_1
&=
\sqrt{\gamma_- t}\, J_-
-\frac{\gamma_-^{3/2} t^{3/2}}{4}
\Bigl(
J_-(J_+J_-)
+
(J_+J_-)J_-
\Bigr),
\label{eq:K1_final}
\\
K_2
&=
\frac{\gamma_- t}{\sqrt{2}}\, J_-^2, \\ 
K_3 & = \sqrt{(I - K_0^{\dagger} K_0  - K_1^{\dagger}K_1 -K_2^{\dagger}K_2)}
\label{app: eq: final set of kraus operators for truncated dynamics},
\end{align}
where \(K_3\) is introduced to satisfy the completeness condition. 

\section{Petz Recovery Map}
\label{appendix: sec: petz_recovery_second_order}
Given the input state $\rho$ and the output state after the error map $\mathcal{E}(\rho)$, the Petz recovery map is defined by the recovery Kraus operators $R_\mu(t)$. These are constructed from the forward Kraus operators $K_\mu(t)$ as:
\begin{equation}
    R_\mu(t) = \rho^{1/2} K_\mu^\dagger(t) \left( \mathcal{E}(\rho) \right)^{-1/2}.
\end{equation}

\subsection{Recovery map for first-order dynamics}

Using the formula the recovery Kraus operators are 
\begin{align}
    R^{(1)}_0(t)
    &=
    \rho^{1/2}
    \left[
    I-\gamma_- t J_+J_-
    -\frac{(\gamma_- t)^2}{4}(J_+J_-)^2
    \right]^{1/2}
    \left[\mathcal{E}(\rho)\right]^{-1/2},
    \label{eq:petz_r0_first_order_full}
    \\[4pt]
    R^{(1)}_1(t)
    &=
    \sqrt{\gamma_- t}\,
    \rho^{1/2}J_+
    \left[\mathcal{E}(\rho)\right]^{-1/2},
    \label{eq:petz_r1_first_order_full}
    \\[4pt]
    R^{(1)}_2(t)
    &=
    \frac{\gamma_- t}{2}\,
    \rho^{1/2}J_+J_-
    \left[\mathcal{E}(\rho)\right]^{-1/2}.
    \label{eq:petz_r2_first_order_full}
\end{align}

If \( \mathcal{E}(\rho) \) is not full rank, the inverse square root is understood as the Moore--Penrose inverse on the support of \( \mathcal{E}(\rho) \).

\subsection{Recovery map for second-order dynamics}
We first compute adjoints
\begin{align}
    K_0^\dagger &= \mathbb{I} - \frac{\gamma_- t}{2} J_+ J_- + \frac{\gamma_-^2 t^2}{8} (J_+ J_-)^2 = K_0, \\
    K_1^\dagger &= \sqrt{\gamma_- t} J_+ - \frac{\gamma_-^{3/2} t^{3/2}}{4} \Big( J_+ J_- J_+ + J_+ J_+ J_- \Big), \\
    K_2^\dagger &= \frac{\gamma_- t}{\sqrt{2}} J_+^2, \\
    K_3^\dagger &= K_3.
\end{align}
Note that $K_3^\dagger = K_3$ holds because the argument inside the square root of $K_3$ is a sum of Hermitian, positive semi-definite operators, making its principal square root Hermitian as well.

Substituting these adjoints back into the Petz recovery formula yields the explicit set of four recovery Kraus operators:
\begin{align}
    R_0 &= \rho^{1/2} \left[ \mathbb{I} - \frac{\gamma_- t}{2} J_+ J_- + \frac{\gamma_-^2 t^2}{8} (J_+ J_-)^2 \right] \left( \mathcal{E}(\rho) \right)^{-1/2}, \\
    R_1 &= \rho^{1/2} \left[ \sqrt{\gamma_- t} J_+ - \frac{\gamma_-^{3/2} t^{3/2}}{4} \Big( J_+ J_- J_+ + J_+ J_+ J_- \Big) \right] \left( \mathcal{E}(\rho) \right)^{-1/2}, \\
    R_2 &= \rho^{1/2} \left[ \frac{\gamma_- t}{\sqrt{2}} J_+^2 \right] \left( \mathcal{E}(\rho) \right)^{-1/2}, \\
    R_3 &= \rho^{1/2} \left[ \sqrt{\mathbb{I} - K_0^\dagger K_0 - K_1^\dagger K_1 - K_2^\dagger K_2} \right] \left( \mathcal{E}(\rho) \right)^{-1/2}.
    \label{eq:app:final recovery map using petz}
\end{align}

\section{Gate complexity of Algorithm 1}
\label{app:gate-complexity of algorithm 1}

Here we derive the primitive gate count used in the main text.  The count is an
upper bound: we do not use cancellations, parallelization, or special structure
of the recovery operators.  We separate the operations into three classes:
single-ancilla gates, system-only unitaries acting on the Dicke subspace, and
controlled system--ancilla Dicke-space unitaries.

The first binary splitting stage has the decomposition
\begin{equation}
U^{(1)}_{S,a}(\Theta_1)
=
\left(e^{i\Theta_1}\otimes \mathbb{I}_a\right)
\left(\mathbb{I}_S\otimes S_a^\dagger\right)
\left(\mathbb{I}_S\otimes H_a\right)
\Lambda(e^{-2i\Theta_1})_{S,a}
\left(\mathbb{I}_S\otimes H_a\right)
\left(\mathbb{I}_S\otimes S_a Z_a\right).
\end{equation}
Thus the first splitting stage requires
\begin{equation}
    N_{\mathrm{1q}}^{(1)}=5,\qquad
    N_S^{(1)}=1,\qquad
    N_{\mathrm{cD}}^{(1)}=1,
\end{equation}
where \(N_{\mathrm{1q}}\), \(N_S\), and \(N_{\mathrm{cD}}\) denote single-ancilla gates,
system-only Dicke-space unitaries, and controlled Dicke-space unitaries,
respectively.  If the ancilla is initialized in \(\ket{0}\), the final \(Z_a\)
gate is redundant for the relevant input branch, reducing the single-ancilla
count to \(N_{\mathrm{1q}}^{(1)}=4\).  We keep the conservative count \(5\) below.

For every subsequent splitting stage, the decomposition has the form
\begin{equation}
U^{(k)}_{S,a_k}(\Theta_k)
=
\left(L_{k0}\otimes \mathbb{I}_{a_k}\right)
\Lambda(L_{k0}^\dagger L_{k1})
U^{(1)}_{S,a_k}(\Theta_k)
\left(F_k^\dagger\otimes \mathbb{I}_{a_k}\right),
\qquad k>1 .
\end{equation}
This contains the first-stage block \(U^{(1)}_{S,a_k}\), two additional
system-only Dicke-space unitaries, \(L_{k0}\) and \(F_k^\dagger\), and one
additional controlled Dicke-space unitary, \(\Lambda(L_{k0}^\dagger L_{k1})\).
Therefore each subsequent split requires
\begin{equation}
    N_{\mathrm{1q}}^{({\mathrm{sub}})}=5,\qquad
    N_S^{({\mathrm{sub}})}=3,\qquad
    N_{\mathrm{cD}}^{({\mathrm{sub}})}=2.
\end{equation}

A rank-\(K\) CPTP map requires \(K-1\) binary splittings: one initial split and
\(K-2\) subsequent splits.  Therefore the total splitting cost is
\begin{align}
    N_{\mathrm{1q}}^{\mathrm{split}}
    &=
    5+5(K-2)
    =
    5(K-1), \\
    N_S^{\mathrm{split}}
    &=
    1+3(K-2)
    =
    3K-5, \\
    N_{\mathrm{cD}}^{\mathrm{split}}
    &=
    1+2(K-2)
    =
    2K-3.
\end{align}
If the redundant \(Z\) gates are omitted, the single-ancilla count becomes
\(N_{\mathrm{1q}}^{\mathrm{split}}=4(K-1)\).  This does not affect the dominant scaling.

The feedback step applies one branch-conditioned unitary for each Kraus
operator.  Counting each feedback as one controlled Dicke-space unitary gives
\(K\) additional controlled operations. Hence the dominant primitive cost of the full coherent recovery circuit overlooking the single-qubit gates is 
\begin{equation}
 N_S=3K-5,
    \qquad
 N_{\mathrm{cD}}=3K-3.
\end{equation}
The total gate count is 
\begin{equation} 
\boxed{ N_{\mathrm{tot}}=N_S+N_{\mathrm{cD}}=6K-8. } 
\end{equation}
Thus the coherent primitive complexity scales linearly in the Kraus rank,
\begin{equation}
   \boxed{ N_{\mathrm{tot}}=O(K)}.
\end{equation}

Finally, let \(g_S(N)\) denote the GPG cost of an arbitrary system-only unitary
on the \(N\)-qubit Dicke subspace, and let \(g_C(N)\) denote the cost of the
corresponding controlled Dicke-space unitary.  The compiled GPG cost satisfies
\begin{equation}
    n_{\mathrm{GPG}}
    \lesssim
    (3K-5)g_S(N)+(3K-3)g_C(N).
\end{equation}
Since from Tab.~\ref{tab: time complexity of elementary operations} Dicke-space synthesis gives \(g_S(N)=O(N^2)\), and the controlled version also has \(g_C(N)=O(N^2)\), we obtain
\begin{equation}
\boxed{
    n_{\mathrm{GPG}}=O(KN^2).
}
\end{equation}
For a generic CPTP map on the Dicke subspace, \(d=N+1\) and \(K\leq d^2\), so
the worst-case compiled complexity is
\begin{equation}
   \boxed{n_{\mathrm{GPG}}=O(N^4)}.
\end{equation}

\end{document}